\documentclass[aps,pre,twocolumn]{revtex4-1}
\usepackage{graphicx}
\usepackage{hyperref}
\usepackage{xcolor}
\usepackage{amsmath}
\usepackage{amsfonts}
\usepackage{amssymb}
\usepackage{varioref}
\usepackage{float}
\usepackage{dcolumn}   % needed for some tables
\usepackage{subfigure}
\newcommand{\cblue}{\color{black}}

\begin{document}
\title{Behavior of active filaments near solid-boundary under linear shear flow}
\author{Shalabh K. Anand }
\email{skanand@iiserb.ac.in}
\author{Sunil P. Singh}
\email{spsingh@iiserb.ac.in}
%\noaffiliation
\affiliation{Department of Physics,\\ Indian Institute of Science Education and Research, \\Bhopal 462 066, Madhya Pradesh, India}
%\date{\today}

\begin{abstract}
The steady-state behavior of a dilute suspension of self-propelled filaments confined between planar walls subjected to the Couette-flow is reported herein. The effect of hydrodynamics has been taken into account using a mesoscale simulation approach. We present a detailed analysis of positional and angular probability distributions of filaments with varying propulsive force and shear-flow. Distribution of centre-of-mass of the filament shows adsorption near the surfaces, which diminishes with the flow. The excess density of  filaments decreases with Weissenberg number as $Wi^{-\beta}$ with an exponent $\beta \approx 0.8$, in the intermediate shear range ($1 < Wi < 30$). The angular orientational moment also decreases near the wall as $Wi^{-\delta}$ with $\delta \approx 1/5$; the variation in orientational moment near the wall is relatively slower than the bulk.  It shows a strong dependence on the propulsive force near the wall, and it varies as $Pe^{-1/3}$ for large $Pe\ge 1$. The active filament shows orientational preference with flow near the surfaces, which splits into upstream and downstream swimming. The population splitting from a unimodal (propulsive force dominated regime) to bimodal phase (shear dominated regime) is  identified in the parameter space of propulsive force and shear flow.
\end{abstract}
\maketitle

\section{Introduction}
 The abundance of swimmers is widespread from microscopic to macroscopic length scales in nature, such as algae\cite{Ringo543,Polin487}, bacteria\cite{Blair1995,Berg1973}, spermatozoa\cite{GRAY775,SHACK1974555,Woolley01082003}, {\it C. elegans}, fishes, etc. The self-propulsion helps them in the endurance of getting food, avoiding various chemical toxins, in reaching to female reproductive egg, and for several other biological processes in complex environments. Typically, motile organisms live near surfaces, which makes them vulnerable to any external perturbations, especially to a  fluid-flow\cite{Tung:softmatt2014,lauga2009hydrodynamics}. The interplay between propulsive force and a flow induced motility is referred to rheotaxis in literature\cite{uspal2015rheotaxis,kantsler2014rheotaxis,rosengarten1988rheotactic}. Understanding  the behavior of self-propelled organisms near a surface is crucial from several bio-applications point of view to a fundamental quest\cite{gao2012cargo}.

 Natural microswimmers play a significant role in various biological process, therefore their dynamics and structure  are subject of immense research interest\cite{Berke:prl:2008,sabass:prl:2010,Schaar:PRL:2015,elgeti2015run,tournus2015flexibility,das2018confined,uspal2015rheotaxis,daddi2018swimming,ledesma2012circle,Elgeti2016,potomkin2017focusing,omori2016upward,tao2010swimming,de2016understanding,pagonabarraga2013structure,zhang:acs:2010,Hill:PRL2007,KAYABiophysical:2012,Yuan:PNAS2015,kantsler2014rheotaxis}. On the other hand, artificially designed microswimmers can be used as a potential model system for targeted delivery in pharmaceutical applications\cite{Howseprl2007,Paxton2006,palacci2015artificial}. Microswimmers' physical behavior can be influenced substantially under external perturbations\cite{Nili:rsc:2017,Chilukuri:jpcm:2014,ezhilan_saintillan_2015}. This can lead them to motile against the stream near  surfaces\cite{Nili:rsc:2017,Chilukuri:jpcm:2014,ezhilan_saintillan_2015,Bretherton490,Zhang:srep:2016,katuri2018cross,rosengarten1988rheotactic,son2015live,rusconi2014bacterial,kaya2009characterization,meng2005upstream}. In living matter, swimming against the flow is common in nature, specifically for the  fishes\cite{Montgomery:nat:1997}, {\it C. elegans}\cite{Yuan:PNAS2015}, {\it E. Coli}, sperms\cite{kantsler2014rheotaxis}, etc. The main difference in the mechanism of swimming at  macroscopic and microscopic  length-scales is the intervention of visual and tactile sensory cues\cite{Montgomery:nat:1997,ARNOLD:1974} in former case, while in the latter case  motion is driven by the interplay of various physical interactions\cite{Marcos:NAS2012}. The physical reason behind the upstream motility is attributed to shear-induced orientation, active stresses, reduction in local viscosity, and inhomogeneous hydrodynamic drag\cite{Berke:prl:2008,Guanglai:PRL2009,lin2000direct}.

 In the literature, simpler yet effective models have been proposed to  unravel dynamics of microswimmers near surfaces\cite{Nili:rsc:2017,Chilukuri:jpcm:2014,ezhilan_saintillan_2015,son2015live,rusconi2014bacterial,kaya2009characterization,meng2005upstream,de2016understanding,najafi2004simple,pande2015forces,babel2016dynamics}. 
  Despite their simplicity, they are able to capture various complex behaviors of microswimmers such as, surface accumulation\cite{Elgeti:2013,Elgeti:2009,Elgeti2016}, upstream swimming\cite{ezhilan_saintillan_2015,Chilukuri:jpcm:2014,Nili:rsc:2017}, and flow-induced angular alignment\cite{Tung:PRL2015,martin2018active}, etc. The population splitting from a unimodal to a bimodal phase is also reported in terms of chirality and  angular speed\cite{Nili2018}. Further, the upstream motion can be regulated  using viscoelastic fluid\cite{Mathijssen:PRL2016}. The propulsion mechanism changes accumulation and angular alignment near  walls, more specifically, puller swimmers orthogonally point towards the wall. Whereas, pusher or neutral swimmers tend to align along the wall\cite{Malgarettijcp:2017}. An external flow has tendency to suppress the excess adsorption of the dimer-like swimmers on the surfaces\cite{Chilukuri:jpcm:2014}.

 In this article, we attempt to provide a thorough study of slender-like motile objects near the solid interface subjected to flow. The influence of flow on the weakening of accumulation and angular distribution near the surfaces is addressed in an elaborate manner. We incorporate  hydrodynamic interactions in our model, which is crucial in the study of active matter systems. The hydrodynamic interactions can induce effective attraction between the wall and an elongated shape swimmer\cite{Elgeti:2009,pedley1992hydrodynamic,berke2008hydrodynamic}. In the previous studies, long-range correlations among solvent, swimmers, and solid-boundaries were not taken into account\cite{Nili:rsc:2017,ezhilan_saintillan_2015}. 
 
 We consider a simulation model that incorporates  an explicit solvent based  mesoscale approach known as multi-particle collision dynamics (MPC)\cite{Malevanets:jcp:1999,Kapral:2008,Gompper2009} clubbed with  molecular dynamics (MD). The dilute suspension of active filaments exhibits an enhancement in average density near a solid-boundary with an increase in P{\'e}clet number, while fluid flow leads to desorption of the swimmers. The density variation of swimmers is quantified in terms of short time diffusion, alignment of a rod-like swimmer across the channel, and residence time near wall and bulk. 
 %The long-range hydrodynamic interactions cause an increase in the adsorption in the weak flow regime.
The swimmers align (anti-align) parallel to the flow at the top (bottom) wall. The shear-induced orientational alignment exhibits a non-monotonic behavior on increasing flow rate, which is attributed to blindness of polarity at higher flow. The majority and minority populations at walls display orientation-switching. In the shear dominant regime, the population splits from a unimodal to bimodal phase. The orientational moment near the surfaces shows  a power law variation on the low strength with an exponent slightly smaller than found in the bulk, it also exhibits a power law scaling with P{\'e}clet number in large $Pe$ regime.
 
 The organisation of the paper is as follows: In section 2, the simulation methodology of the self-propelled filaments and  the fluid has been discussed. Results are presented in section 3, with the discussion  of the competition among the flow, confinement, and active forces. We have summarised our study in the section 4.

\section{Model}
\begin{figure}%[h]
 	\includegraphics[width=\linewidth]{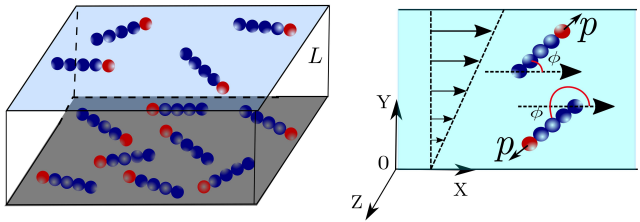}
 	\caption{A schematic picture of active filaments confined between walls. Bottom wall is in grey and top wall is shown in light blue for the better visibility. A red bead shows the head of swimmers, thus their direction of polarity. Arrow in the right diagram displays direction of flow.}
 	\label{Fig:model}
 \end{figure}

In this section, we present simulation method adopted for active filaments in solution. At first, modelling of an active filament is presented, and subsequently implementation of a coarse-grained model for the  solvent  is introduced. A schematic display of the  system confined along y-direction is shown in Fig.~\ref{Fig:model}. In other two spatial directions (x and z), periodic boundary condition is applied.  
  
\subsection{Active Filament}
 We consider $N_p$ filaments, where each filament consists a linear sequence of $N$ monomeric units, connected via spring and bending potentials.
 Thus total  $N_t=N
 \times N_p$ number of monomers are present  in the solution.  
  An excluded volume interaction among the monomers and walls are also taken into account.  The total potential energy of  a filament is written as 
 $U = U_{sp} + U_{b} + U_{LJ}+U_{w}$.
 Here $U_{sp}$, $U_b$, $U_{w}$, and $U_{LJ}$ are harmonic, bending, wall and repulsive part of the Lennard-Jones (LJ) potentials, respectively. The harmonic and bending potentials for $j^{th}$ filament is given as, 
\begin{equation}
	U^{j}_{sp} + U^{j}_{b}= \frac{k_{s}}{2} \sum_{i=1}^{N-1}(|{\bf R}_{i}|-l_{0})^2 +  \frac{\kappa}{2} \sum_{i=1}^{N-2}({\bf R}_{i+1}- {\bf R}_{i})^2,
	\label{bond_bend}
\end{equation}
 where $l_0$ is the equilibrium bond length, $R_i$ is the length of the $i^{th}$ bond vector, $R_i= |{\bf r}_{i+1}-{\bf r}_{i}|$ with ${\bf r}_i$ to be position vector of the $i^{th}$ monomer, $k_s$ and $\kappa$ are spring constant and bending rigidity of the filament, respectively.

The excluded volume potential $U_{LJ}$ is modelled as repulsive part of LJ potential for shorter distance, i.e., $R_{ij} < 2^{1/6}\sigma$, among all monomers,
\begin{equation}
	U_{LJ} = \sum_{i=1}^{N_t-1} \sum_{j=i+1}^{N_t} 4 \epsilon  \left[\left(\frac{\sigma}{R_{ij}}\right)^{12}- \left( \frac{\sigma}{R_{ij}}\right)^{6} + \frac{1}{4} \right], 
	\label{Eq:LJ}
\end{equation}
 and for $R_{ij} \geq 2^{1/6}\sigma$, $U_{LJ}$ is treated to be zero. Here, $\epsilon$ and $\sigma$ are the LJ interaction energy and the diameter, respectively. Interactions between wall and monomers ($U_{w}$) are also treated in same manner as given in Eq.~\ref{Eq:LJ}, to constrain the filaments within wall premises. A monomer feels the repulsive force from a boundary wall when it reaches within the distance $2^{1/6}\sigma/2$ from the wall.

The self-propulsion is achieved by imposing tangential force along each bond vector of the filament, thus force  on $i^{th}$ filament can be written as  $F_a^{i}= \sum_{j=1}^{N-1}f_a {\hat t}({\bf R}_j) $\cite{isele2015self,anand2018structure}, $f_a$ and ${\hat t}({\bf R}_j)$ reads as the strength of the active force, and  $j^{th}$ tangent vector, respectively. 

\subsection{MPC fluid}
 The MPC method is also known as stochastic rotation dynamics approach\cite{Malevanets:jcp:1999,Kapral:jcp2000,Kapral:2008,Gompper2009,Ihle:PRE:2001}, where solvent molecules are treated as point particles of mass $m$. Their dynamics consist of streaming followed by collision in the alternating steps. In the streaming step, solvent particles move ballistically with their respective velocities, and their positions are updated according to the following rule, ${\bf r}_{i}(t+h) = {\bf r}_{i}(t) + h {\bf v}_{i}(t)$, where $h$ is the MPC collision time-step and $i$ is the index for a solvent molecule. In the collision step, solvent molecules are sorted into cubic cells of side $a$ and their relative velocities, concerning the centre-of-mass velocity of the cell, are rotated around a randomly oriented axis by an angle $\alpha$. The particles' velocities are updated as
	
	\begin{equation}
	{\bf v}_{i}(t+h) = {\bf v}_{cm}(t) + \Omega(\alpha)({\bf v}_{i}(t) - {\bf v}_{cm}),
	\label{collision}
\end{equation}
 where ${\bf v}_{cm}$ is the centre-of-mass velocity of the cell of $i^{th}$ particle, and $\Omega(\alpha)$ stands for the rotation operator. During collision, all solvent molecules within a cell interact with each other in a  coarse-grained fashion by colliding at the same time to ensure the momentum conservation. This ensures the long-range spatial and temporal correlations among the solvent molecules that results hydrodynamic interactions.

{\cblue  A solvent molecule's velocity is reversed by bounce-back rule as ${\bf v}_i=-{\bf v}_i$\cite{lamura2001multi,lamura2002numerical,Singh_2014_JCP}, when it collides with a wall during streaming step. This imposes no-slip boundary conditions on both walls}. The interaction of MPC-fluid and filament monomers are incorporated during the collision step. Here, momentum of monomers in the calculation of centre-of-mass velocity  of the cell is included during the collision step\cite{Malevanets:EPL:2000,Ripoll:EPL:2004}.

Furthermore, an active force on the filament adds momentum in the polarity direction of a filament, which destroys the local momentum conservation. To insure local momentum conservation, the same force to the solvent particles are imposed in opposite direction  to only those cells where monomers are present during each collision step\cite{Elgeti:2009}. 
The presence of propulsive force and flow continuously increases energy of the system, which may lead to rise in the temperature of fluid. A cell level canonical thermostat known as Maxwell-Boltzmann scaling\cite{CCHuang2010,CCHuang2015} is incorporated  to remove the excess energy and keep the desired temperature of the system. 
 A random shift of the collision cell at every step is also performed to avoid the violation of Galilean invariance\cite{Ihle:PRE:2001,kroll:pre2003}.

 A linear fluid-velocity ($v_{x} = \dot{\gamma} y$) profile along the x-axis is generated by moving the wall at $y=L_y$ (top wall in the Fig.\ref{Fig:model}) at constant speed ($v_{x} = \dot{\gamma} L_y$). {\cblue This gives a flow profile as shown in Fig.~\ref{Fig:model}, and it has a net flow along x- direction. } The equations of motion of solvent molecules are modified in the  vicinity of wall\cite{Winkler:jcp:2009,Whitmer:2010,Singh_2014_JCP}. The velocity of a solvent particle relative to the surface is reversed with bounce-back rule in the streaming step when it touches a wall, which again ensures no-slip boundary conditions on the walls. During the solvent-wall interaction, a moving wall transfers momentum to the solvent molecules which drives a linear profile on average along the  x-direction. In addition to that, virtual particles with the velocity taken from a Maxwell-Boltzmann distribution with mean equal to the wall velocity are added in the partially filled cells\cite{CCHuang2010,CCHuang2015}.
 
\subsection{Simulation Parameters} 
 All the physical parameters are  scaled in terms of the MPC cell length $a$, mass of a fluid particle $m$, thermal energy $k_{B}T$ and time $\tau=\sqrt{ma^2/k_BT}$. The size of simulation box are taken here as ($L_x=80,L_y= 50,L_z= 25$)  with periodic boundary conditions in x and z spatial directions, and solid walls are in y-directions at $0$ and $L_y$. Other parameters  are chosen as spring constant $k_{s}=1000k_{B}T/l_{0}^{2}$, stiffness parameter $\kappa=5000k_{B}T/l_{0}^{2}$, $l_{0}/a =\sigma/a=1$ and $\epsilon/k_{B}T=1$. Unless explicitly mentioned, $N=10$ (number of monomers in a filament)  and   $N_p=50$ (number of filaments), which results a dilute concentration of monomer in solution as $\rho_m=0.005a^{-3}$, and number density of rod is $\rho_p=0.0005a^{-3}$. This  is well below the isotropic-nematic transition\cite{prost1995physics}. The velocity-Verlet algorithm is used for the integration of equations of motion for  active polymers, and integration time-step is chosen to be $h_m=0.01\tau$. The strength of active force is measured in terms of a dimensionless quantity called  P{\`e}clet number, which is defined as $Pe = \frac{f_{a} \sigma}{k_{B} T}$.
 
 For the MPC fluid, collision time-step is taken as $h = 0.05\tau$, rotation angle $\alpha=130^{\circ}$, average number of fluid particles per cell $<N_{c}>=10$. These parameters correspond to the transport coefficient of the fluid as zero-shear viscosity $\eta_s \equiv 17 \sqrt{m k_{B}T}/a^{2}$. The strength of flow is expressed in terms of a dimensionless Weissenberg number $Wi$ defined as $Wi = \dot{\gamma} \tau_r$, where $\tau_r$ is the polymer's relaxation time. Here, $Wi$ is a measure of the flow strength over the thermal fluctuations, in the limit of $Wi\le 1$, thermal fluctuations dominate, however for $Wi\ge 1$, flow plays a significant role. All the simulations are performed in the range of $0 \le Wi \le 150$ and $0\le Pe \le 5$. For each data set, 50 independent runs have been generated for better statistics. All the simulations results are below the Reynolds number $Re<0.1$. 
 
\section{Results}
 In our model, an active filament  moves  along one of its end, thus it is a polar filament. It can align along the surfaces, which leads to absorption on the surfaces and depletion in  bulk\cite{Elgeti:2013,Elgeti2016}.  A polar active filament may preferentially align  parallel or anti-parallel to the flow direction. Here, we unravel the influence of linear shear-flow on a dilute suspension of active filaments in a confined channel especially, on surface adsorption, average local orientation profile, population of upstream swimmers near the wall and also the importance of hydrodynamic interactions on swimmers.
 
 \subsection{Surface accumulation}
\begin{figure}
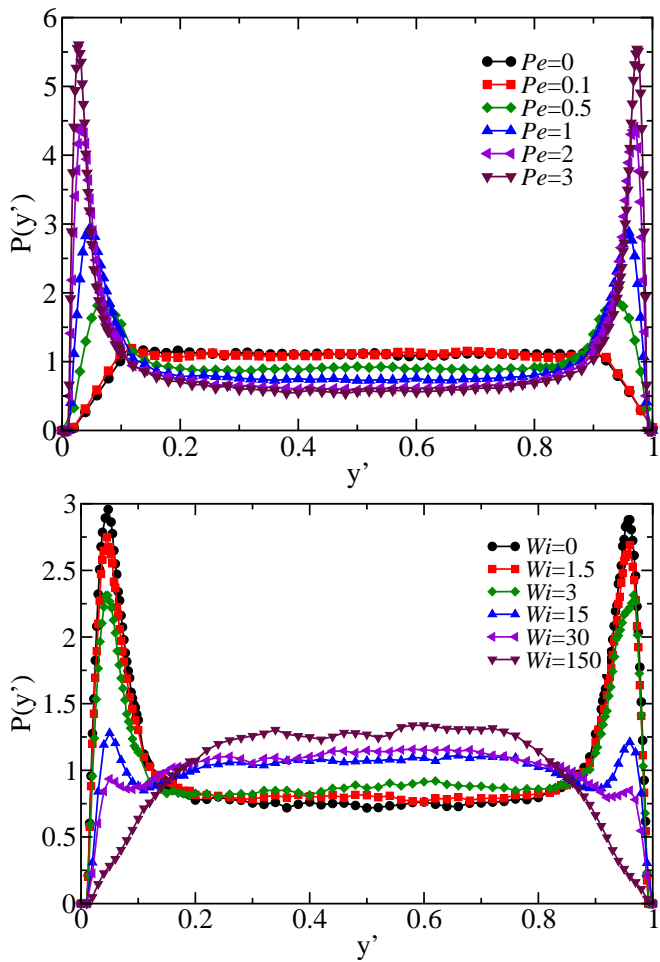
%[h]
	\includegraphics*[width=\linewidth]{accum_wi0}
	\includegraphics*[width=\linewidth]{accum_dist}
	\caption{The probability distribution of active filaments as a function of channel height ($y'=y/L_y$), Fig. a) shows for $Wi=0$ for various $Pe$, and Fig. b) displays for various flow-rates at $Pe=1$.  }
	\label{Fig:accumulation}
\end{figure}

The distribution of passive filaments is shown in Fig.~\ref{Fig:accumulation}-a at $Wi=0$ as a function of distance from the bottom wall $y'=y/L_y$, where walls are present at $y'=0$ (bottom) and $y'=1$ (top). The probability distribution $P(y')$ is uniform due to translational entropy, which favours homogeneity. Near the surfaces ($|L_y-y|\le l/2$), it reflects depletion due to steric repulsion of the filaments from the wall. As expected, $P(y')$  of the active filament increases with the speed of  swimmers (see Fig.\ref{Fig:accumulation}-a) near the surfaces. The normalised distribution of swimmers near the surfaces reflects large inhomogeneity especially in the limit of large $Pe$. This is consistent with the previous findings\cite{Elgeti:2013,Elgeti2016,Chilukuri:jpcm:2014}. The distribution function has two identical peaks near the walls in the limit of large P{\'e}clet number, which reflects the adsorption of swimmers on the solid boundaries. 

 The motility induced adsorption is understood in terms of the combination of active force, steric repulsion, and viscous-drag. Large active force results in longer persistence motion and rapid movement throughout the channel, which eventually causes filaments to reach on  surfaces at a shorter time interval. Once they reach nearby the surface, they align and move along the surface. {\cblue The relatively larger drag perpendicular to  filament's axis causes it to reside  near the wall for a longer time compared to the bulk, which results in higher probability density with $Pe$. This will be elucidated in terms of short time diffusion across the channel. However, inhomogeneous drag is not the sole reason for surface adsorption. Width of the channel, length of the filament, and its rotational diffusion coefficient also affect the adsorption significantly\cite{li2009accumulation,elgeti2013wall}.  }

  \begin{figure}%[h]
	\includegraphics*[width=\linewidth]{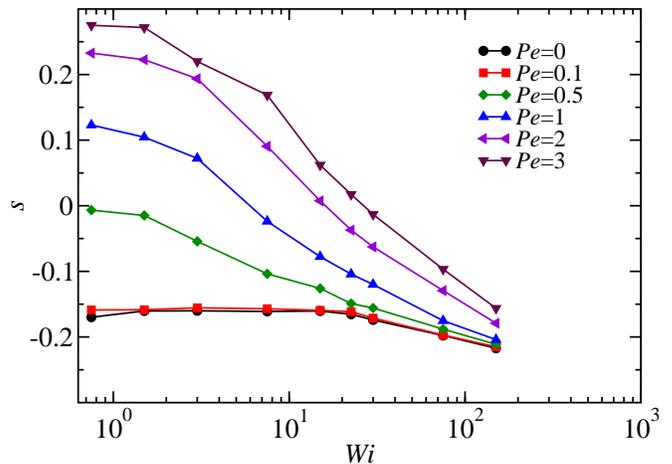}
	\caption{Surface excess as a function of flow-rate $Wi$ for various $Pe$ as shown in figure.}
	\label{Fig:surface_excess}
  \end{figure}
 
 %  \begin{figure}%[H]
 %	\includegraphics*[width=\linewidth]{average_resid.eps}
 %	 	\caption{Average residence time of the active filament near the planner surfaces  as function of $Wi=0$ for various $Pe$  as shown in figure.}
 %	\label{av_residence}
 %\end{figure}

 The flow disrupts the concentration of filaments on the surfaces, which is reflected in probability density in Fig.~\ref{Fig:accumulation}-b for a range of $Wi$ at a fixed $Pe=1$. The surface adsorption is maximum for $Wi=0$ (without flow) at a given $Pe$. The height of peak slowly diminishes with the increase in  fluid-flow for $Wi>1$, and it disappears nearly at $Wi\sim 30$. In the limit of $Wi>100$, density of rods near the surface is very small compared to bulk density. Interestingly, this is quite similar to the  equilibrium distribution (see Fig.~\ref{Fig:accumulation}-a for $Pe=0$ at $Wi=0$).

 The desorption driven by flow can be justified in terms of  orientational alignment of the polar filament, tumbling motion, and suppression of diffusion across the channel. An active filament aligns along the flow direction, which breaks the rotational symmetry of swimmers causing less number of filaments aligned towards a wall. Consequently, the probability of a filament residing near the wall decreases. Thus, the accumulation close to the wall is severely influenced by the shear-induced orientational alignment. Therefore, flow can act as a control variable for  adsorption of the polar filaments at the wall.
  
 To quantify the depletion, we estimate surface excess from the probability distribution of the filament as\cite{Elgeti:2009}
 \begin{equation}
 	s = \int_{0}^{l/2}[P(y)-P_{b}]dy,
 	\label{Eq:surf_ex}
 \end{equation}
  where $s$ is a measure of excess surface density near the surface relative to bulk.  Here, $P_b$ is read as bulk distribution. The definition of $s$ is constructed in a way that it becomes unity for full adsorption, whereas in case of uniform distribution it is zero. Smaller density relative to bulk results in a negative surface excess as that  of the passive filament.
    
 For slow swimming speed $Pe<1$, probability density at wall is smaller than the bulk, thus the surface excess is negative. It increases with $Pe$ and becomes positive for $Pe>1$ (see Fig.~\ref{Fig:surface_excess}). The variation in $s$ with flow strength $Wi$ in the limit of $Pe<1$ is negligible, there is significant change in $s$ for larger P{\'e}clet number $Pe>1$ as  Fig.~\ref{Fig:surface_excess} illustrates. The change in $s$ grows with $Pe$ as a function of {\cblue Weissenberg} number. At a sufficiently high value of $Wi$, surface excess becomes negative suggesting the weakening of localisation of filaments at the walls. Desorption of filaments in the large $Wi$ attributes to the dominance of shear over active forces.

 The qualitative behavior of a surface adsorption can also be accessed from the scaling arguments. In this approach, trajectory of a swimmer is treated as a semi-flexible polymer\cite{Elgeti:2009} with persistence length of trajectory defined as $\xi_{p} = v/D_{r}$, where $v$ and $D_{r}$ are the average speed of the filament and the rotational diffusion coefficient, respectively. Furthermore, under a linear shear-flow, rotational diffusion can be approximated in the form of tumbling time as\cite{winkler2006semiflexible,Huang:Macromol2010}
 \begin{equation}
 D_{r} \approx \frac{k_{B}T}{\eta_s l^{3}}(1+c Wi^{-2/3})^{-1},
 \end{equation}
  where $c$ is some constant. We present analysis in two extreme limits. In the limit of $Wi<<1$, $D_{r}$ is unperturbed by fluid-flow, whereas for $Wi>>1$, $D_{r}$ varies as $Wi^{-2/3}$. 
  
 The probability of finding a filament in the vicinity of thickness $l/2$ from a wall is given as $p = \tau_{w}/(\tau_{w}+\tau_{b}) = 1/(1+\tau_{b}/\tau_{w})$, with $\tau_{w}$  to be the residence time of a filament at the wall and $\tau_{b}$ to be the residence time in bulk. In order to estimate $\tau_w$, we take $x_f$ to be the distance travelled along the surface. It is simply $x_f=vt$ and $y$ is normal distance of filament from the closest wall, then $y=x_f^{3/2}/\xi_{p}^{1/2}$ and at $y=l/2$, time is $t=\tau_{w}$. Following the same approach as derived in Ref.~\cite{Elgeti:2009} for the active filaments, we have extended the scaling approach under linear shear-flow. Therefore, the residence time $\tau_{w}$ can be expressed as, 
 \begin{equation}
 \tau_{w} \sim \frac{1}{v} \Big(\frac{l^{2}\xi_{p}}{4}\Big)^{1/3} = \Big(\frac{l^{2}v^{-2}}{4D_{r}}\Big)^{1/3} .
 \label{Eq:tau_w}
 \end{equation}
 Here, the persistence length $\xi_{p}>L_y/2$ for all $Pe$, thus we approximate  $\tau_{b}$ in diffusive regime as\cite{Elgeti:2009},
 \begin{equation}
 \tau_{b} \sim \frac{L_{y}}{v_{b}} \Big(\frac{l}{\xi_{p}}\Big)^{1/3} = \frac{L_{y}}{v_{b}} \Big(\frac{lD_{r}}{v}\Big)^{1/3},
 \label{Eq:tau_b}
 \end{equation}
 with $v=Pe/\gamma+a_{2}Wi$ and $v_{b}=Pe/\gamma$, where $\gamma$ is friction coefficient. Combining expression \ref{Eq:tau_w} and \ref{Eq:tau_b},
 
 \begin{equation}
 \tau_{w}/\tau_{b} \sim \frac{v_{b}}{L_{y}} \Big( \frac{lv^{-1}}{4D_{r}^{2}} \Big)^{1/3}.
 \label{Eq:ratio}
 \end{equation}
 In the limit of $Wi<<1$, the ratio varies as $\tau_{w}/\tau_{b} \sim v^{2/3}$, which recovers the results proposed by {Elgeti} {\it et al}\cite{Elgeti:2009}. In the limit of $Wi>>1$, $\tau_{w}/\tau_{b} \sim Wi^{-\beta}$ with $\beta=7/9$, hence the probability of finding a filament near the wall also becomes $p \sim Wi^{-\beta}$. One can approximate surface excess in the large shear limit as,
  \begin{equation}
 s = \frac{p L_y - l}{L_y-l} \sim \frac{a_{0}Wi^{-\beta} - l}{L_{y}-l}.
 \label{Eq:surf_final}
 \end{equation}
 Furthermore, $Wi^{-\beta} \to 0$ for $Wi >>1$, which leads to the saturation of surface excess at $s=-l/(L_{y}-l)$. In the intermediate limit $Wi>1$, $s$ decrease as a power law $s\sim Wi^{-\beta}$ with an exponent $\beta \sim 0.8$, as displayed in Fig.~\ref{Fig:surface_excess}. 
 
 \subsection{Residence time and Diffusion }
 \begin{figure}
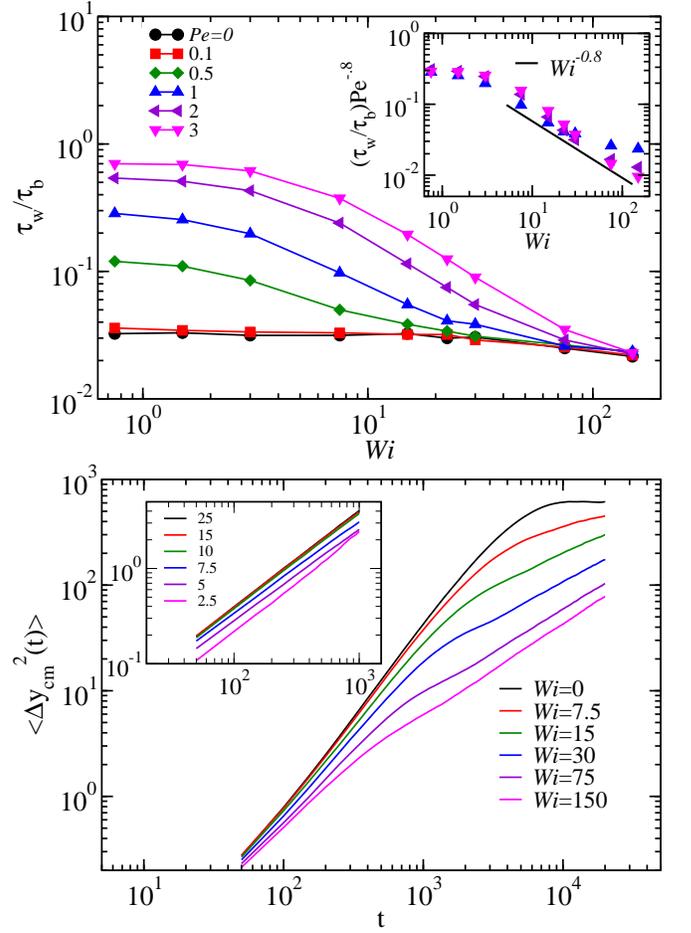
%[h]
 	\includegraphics*[width=\linewidth]{ratio_wall_bulk}
 	\includegraphics*[width=\linewidth]{msd_pe2}
 	\caption{a) Ratio  of residence time of the active filament near the surface with the bulk as a function of $Wi$ for various $Pe$. Inset shows the scaled curve of $(\tau_b/\tau_w )Pe^{-0.8}$ in range of $Pe\ge 1$. b) The MSD of  centre-of-mass of filament for various $Wi$ at a given $Pe=2$. Inset shows the  MSD  as function of time for various distances from the channel.}
 	\label{Fig:msd}
 \end{figure}
 To clarify  effect of shear on the surface adsorption, we estimate residence time of the filament near the wall and in bulk. The time spent by a filament in the neighborhood of  wall with a constraint that once it aligns along the wall is defined as residence time $\tau_w$. Similarly, average time spent in the bulk is defined as twice of the time taken by a filament to reach  either of the surfaces from the centre. The ratio of residence times is displayed in Fig.~\ref{Fig:msd}-a as a function of $Wi$ for various $Pe$. The qualitative  behavior of $\tau_w/\tau_b$ is similar  as surface excess (see Fig.\ref{Fig:surface_excess}). The ratio decreases with  flow and increases with $Pe$. Our simulation results follow the relation $\tau_{w}/\tau_{b} \sim Pe^{0.8}$ for small $Wi$. The exponent is slightly larger than the scaling predictions obtained in Eq.~\ref{Eq:ratio}. The influence of shear is also displayed here as $\tau_{w}/\tau_{b} \sim Wi^{-\beta}$ with $\beta \approx 0.8$ in the intermediate limit of $Wi$ for $Pe>1$. The same exponent is also obtained from scaling behavior in Eq.~\ref{Eq:ratio}. The weak flow has negligible influence on $\tau_w/\tau_b$. On the other hand, stronger flow (for $Pe>1$) can lead to large change in $\tau_w/\tau_b$ (see Fig.~\ref{Fig:msd}-a). The influence of active forces is weak in the case of $Pe<1$, therefore shear has negligible influence on $\tau_w/\tau_b$, thus the adsorption is independent of shear flow. This also establishes the influence of activity and external flow on the accumulation of swimmers.
 
 Furthermore, in the limit of $Wi>>1$, $\tau_w/\tau_b$ saturates to a slightly smaller number than the passive case. This is associated with the alignment of filaments along the flow leading to transport via transverse diffusion across the channel. Hence, $\tau_w/\tau_b$ is constant irrespective  of flow and propulsive forces in this regime. 
 
 We quantify the influence of surface and flow on the short time diffusion across the channel. The  mean-square-displacement(MSD) of the centre-of-mass of the filament along the gradient direction at $Pe=2$ for various $Wi$ is displayed in Fig.~\ref{Fig:msd}-b. For $Wi=0$, the MSD shows super-diffusive regime in short time succeeded by saturation in long time limit. The saturation occurs due to presence of wall, which is nearly at half-width of the channel. The flow strength of $Wi>1$ aligns filament along x-axis, thus the super-diffusive regime appears relatively at shorter time and for narrow window. The diffusive behavior appears at shorter time scales as shown in Fig.~\ref{Fig:msd}-b, and diffusion across the channel is  much slower with $Wi$. The decrease in the short time diffusion  across the channel at sufficiently higher shear-rate explains the lower density of the filament on the surfaces. The flow aligns them thus suppresses the ballistic motion of filaments in the gradient direction, consequently leading to a reduction in density on both surfaces.

 Hydrodynamic interactions influence the diffusion nearby surfaces. This is assessed more intricately in terms of the short time MSD across the channel with separation from the surface. Inset of Fig.\ref{Fig:msd}-b displays the MSD for various distances from the wall at $Pe=0$ and $Wi=0$. The short time MSD is nearly independent in bulk. Interestingly, it shows strong variation in the vicinity of the surface compared to that in bulk. The decrease in the short time diffusion occurs due to alignment along the surface, which forces filament to diffuse  perpendicular to their axis. This exhibits a higher drag and leads to smaller MSD. The inhomogeneous drag and hydrodynamic interactions contribute to slow translational and rotational diffusion.
  
 Despite the decrease in absolute value of residence time\cite{Elgeti:2009}, surface adsorption grows with propulsive force. This is addressed here as follows, if a filament's polarity in bulk is pointing upwards then it may reach to the top wall and if it is aligned downward then it may end up to the bottom wall. The time required to reach on a surface drastically reduces with propulsive force, results in enhancement of the probability of collision with walls. This is clearly demonstrated in Fig.~\ref{Fig:msd}-a as ratio of $\tau_w/\tau_b$. Therefore, adsorption on the surfaces increases even though the residence time  $\tau_w$  decreases with $Pe$. Thus, the decrease in surface excess is also attributed  in terms of MSD across the channel in  Fig.~\ref{Fig:msd}-b. We can conclude that the flow diminishes the persistence motion across the channel, which can be enhanced with the larger propulsive force.
 
  \begin{figure}
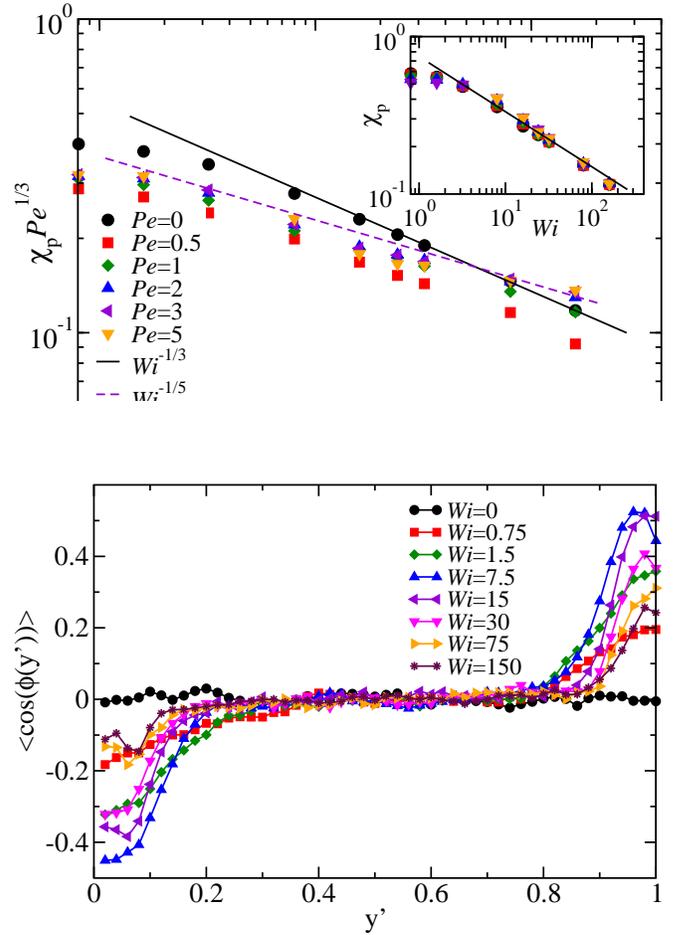
%[h]
  	\includegraphics[width=\columnwidth]{alignment_inset}
  	\includegraphics[width=\linewidth]{orientpe05}
  	\caption{a) Orientation moment  $\chi_p$ in the vicinity of the wall, it  follows  power law relation as $Wi^{-1/5}$ (dashed line) for $Pe>1$ and solid line shows. $Wi^{-1/3}$ for $Pe=0$. Inset shows a universal curve for the $\chi_p$ in the bulk for all $Pe$. Solid line shows power law variation with exponent $1/3$. 	
  		b)Average orientation profile of active filaments as a function of channel height $(y'=y/L_y)$ from the bottom wall for $Pe=0.5$ for various $Wi$.}
  	\label{Fig:orient}
  \end{figure}
  
  \begin{figure*}%[h]
  	\includegraphics*[width=0.2\linewidth]{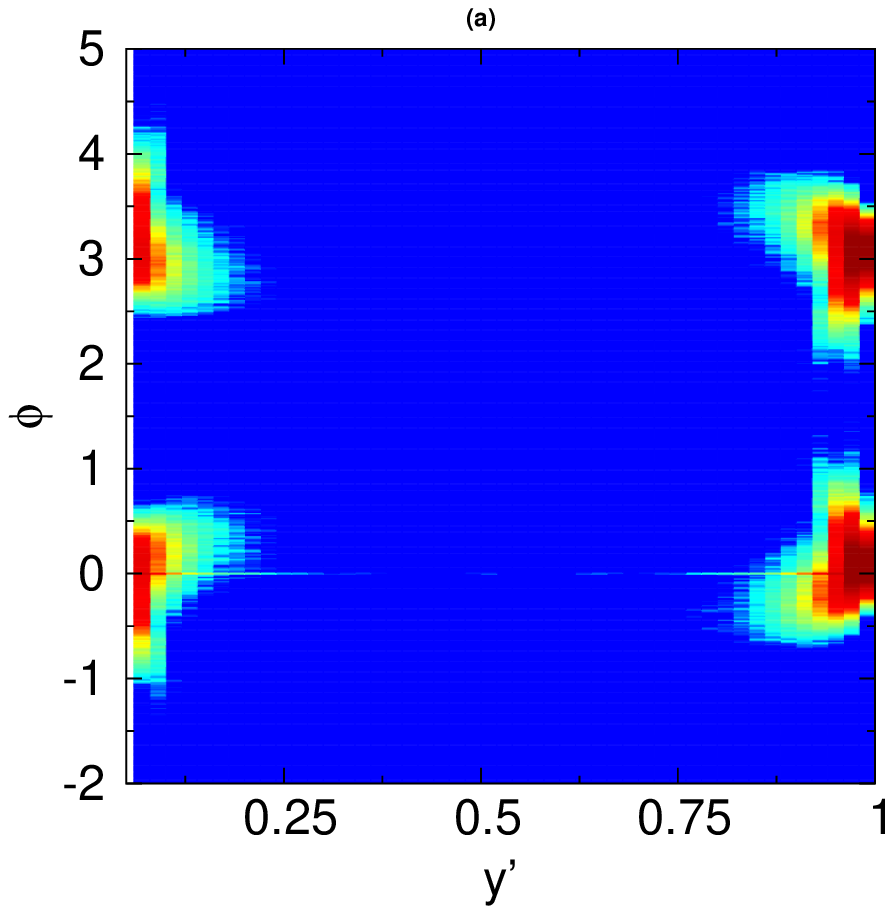}%
  	\includegraphics*[width=0.2\linewidth]{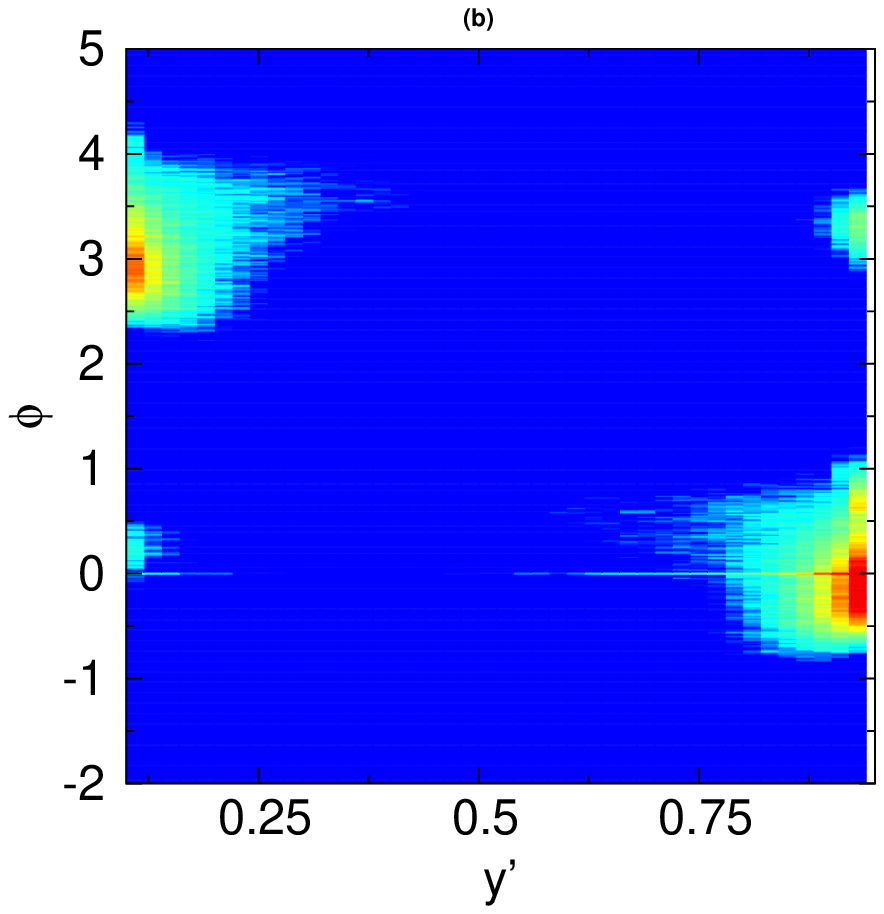}%
  	\includegraphics*[width=0.2\linewidth]{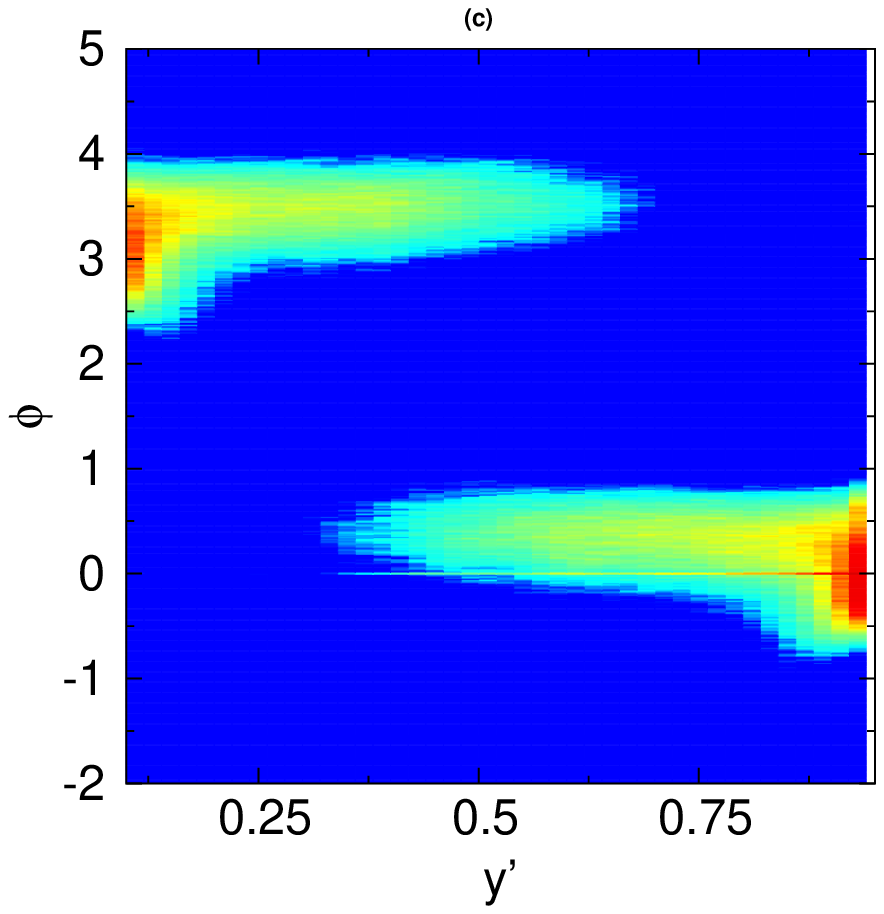}%
  	\includegraphics*[width=0.2\linewidth]{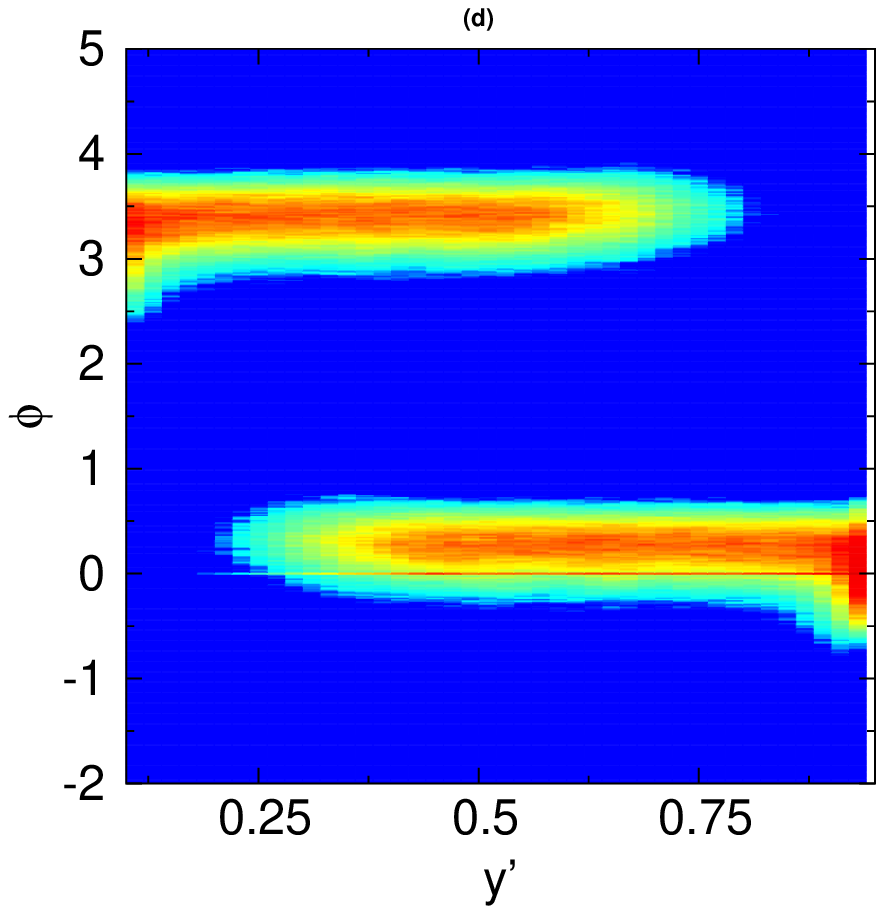}%
  	\includegraphics*[width=0.2\linewidth]{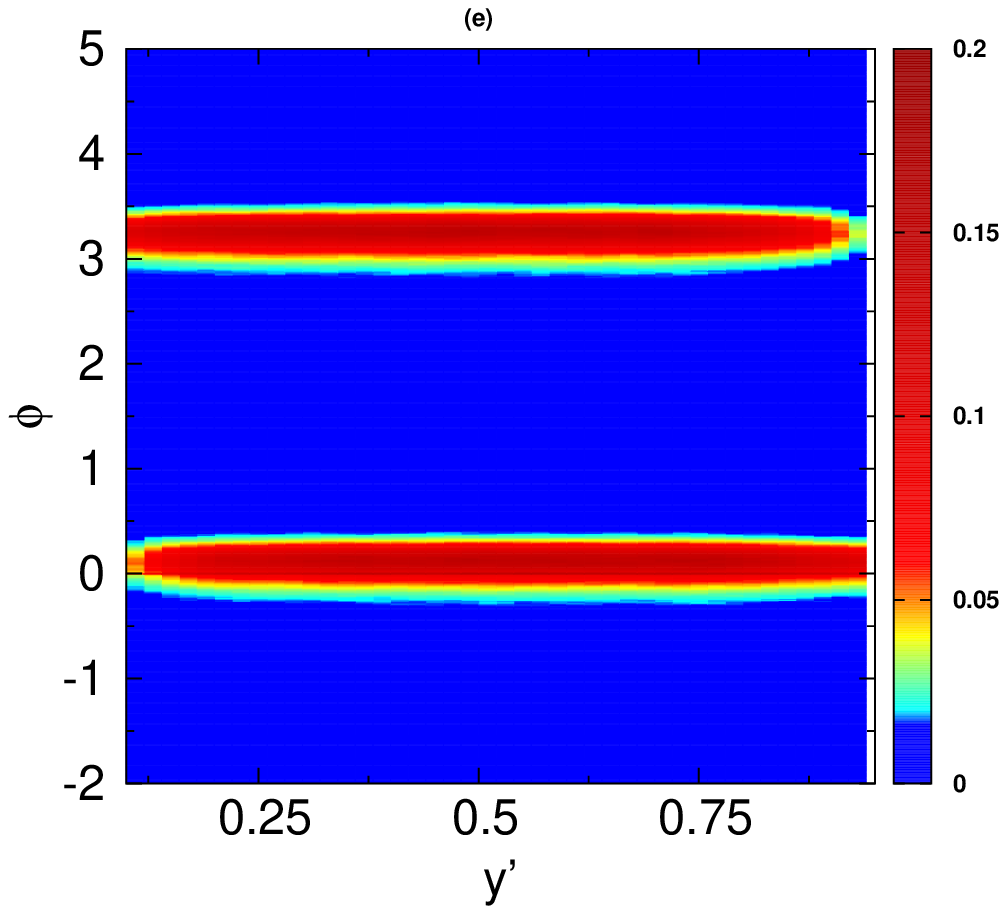}%
  	\caption{The distribution of angle $\phi$ made by the filaments as a function of channel width for a) $Wi=0$, b) $Wi=3$, c) $Wi=7.5$, d)$Wi=15$, and e) $Wi=150$ at $Pe=3$.}
  	\label{Fig:orient:heat}
  \end{figure*}
 
\subsection{Flow Induced Alignment}
  Now we embark on the quantification of flow-induced orientation of active filaments particularly near the surfaces, which enables  characterization of direction of a swimmer. For this, we compute two different angles, one is the angle between flow direction and the projection of unit vector $\textbf{p}$ (tail-to-head as shown in Fig.~\ref{Fig:model}) on the flow-gradient plane, and latter one is the angle between $\textbf{p}$ and flow direction. The former one is azimuthal angle used in  characterisation of the shear-induced alignment of a polymer\cite{Huang:Macromol2010,Chien-Cheng:2012}, and the latter one is used in identifying upstream and downstream swimming behavior\cite{Nili:rsc:2017,Chilukuri:jpcm:2014}.
   
 The variation of orientational moment along the flow direction, defined as $\chi_p=1-{ p}_x {p}_x$, as a function of $Wi$ and $Pe$ in bulk (inset) and close to the wall is displayed in Fig.\ref{Fig:orient}-a. Swimmers are oriented randomly in all possible ways in bulk at low $Wi$, which align along the flow direction for higher $Wi$. The variation of the alignment in the bulk shows a power law behavior as $Wi^{-\delta}$ with  $\delta=1/3$\cite{park2009inhomogeneous,rahnama1995effect,chen1996rheology,leal1971effect}. Note that the scaling exponent is nearly independent of propulsive force. A qualitative picture is shown from a solid line in inset of Fig.~\ref{Fig:orient}-a. In the neighborhood of solid-boundaries, swimmers are more aligned along the flow due to propulsive and excluded volume forces. The angle decreases with increase in $Pe$, thus $\chi_p$ shows slower variation with $Wi$ for large $Pe\ge 1$. This is also reflected in the scaling exponent as $\delta\sim 1/5$ for all $Pe \ge 1$. Interestingly, the angular alignment also shows a power law with $Pe$ near the wall. A universal curve for all $Pe \ge 1$ is obtained by scaling with $Pe^{1/3}$ (see Fig\ref{Fig:orient}-a). This suggests that the orientational moment decreases as $\chi_p \sim Pe^{-1/3}$ for all $Wi$. In the large shear limit, it varies as $\chi_p \sim Wi^{-1/5}$ for $Pe\ge 1$. 
 
 Now we compute the angle between vector $\bf p$ (tail-to-head) and flow direction as shown in  Fig.\ref{Fig:model}. In our convention, if a filament's orientation angle lies in the range of $-\pi/2 \le \phi\le \pi/2$ then it is said to be a downstream swimmer, similarly, if this is between $\pi/2 \le \phi \le 3\pi/2$ then it is referred to an upstream swimmer. The profile for various $Wi$ ( see Fig.\ref{Fig:orient}-b) is shown across the channel for $Pe=0.5$. In absence of flow ($Wi=0$), a filament does not have any preferred direction, thus $<\cos(\phi(y'))>\approx 0$ throughout the channel. This suggests that the swimmers move symmetrically in all directions for $Wi=0$.
 
 For a non-zero $Wi$, symmetry is broken therefore $<\cos(\phi(y'))>$  displays  a non-zero value. A positive value corresponds to downstream swimming in the upper halves ($y'\geq 0.5$) and a negative value for upstream in the lower halves ($y'\leq 0.5$). The average alignment of the system grows with the flow, as Fig.~\ref{Fig:orient}-b exhibits a peak near the top wall and a depth near the bottom wall. The height of peak grows with $Wi$, after reaching  to a maximum value it start to decline (see Fig.~\ref{Fig:orient}-b for $Wi>30$ at $Pe=0.5$). In addition to decrease in the height of  peak, the width of the distribution also gets narrower. This illustrates localisation of the upstream and downstream swimming in the vicinity of the surfaces. {\cblue On the bottom  surface filaments move against the flow, thus the average net flow is upstream  in the intermediate shear range. However in the range of $Wi>>1$, average net flow becomes downstream. Similarly on the top wall, net flow is always downstream.}   
 
 The average preference of swimmers  diminishes as we go far from the surfaces. This can be visualised in terms of distribution of angle $\phi$ in the shear-gradient plane. Figure~\ref{Fig:orient:heat} displays a colormap of angle distribution $P(\phi)$ along  the channel $y'=y/L_y$ for various values of $Wi$ at $Pe=3$. It exhibits two symmetric peaks at $\phi=0$ and $\pi$, which correspond to $Wi=0$. With flow, one of the peaks diminishes on both walls leaving only single dominant phase as displayed in  Fig.\ref{Fig:orient:heat}-c and d. In the limit of large $Wi$, both halves exhibit symmetric distribution around $0$ and $\pi$, thus  nearly equal concentration of upstream and downstream swimmers are present.
 
  \begin{figure}%[h]
  	\includegraphics*[width=\linewidth]{bimodal_topwall_pe05}
  	\includegraphics*[width=\linewidth]{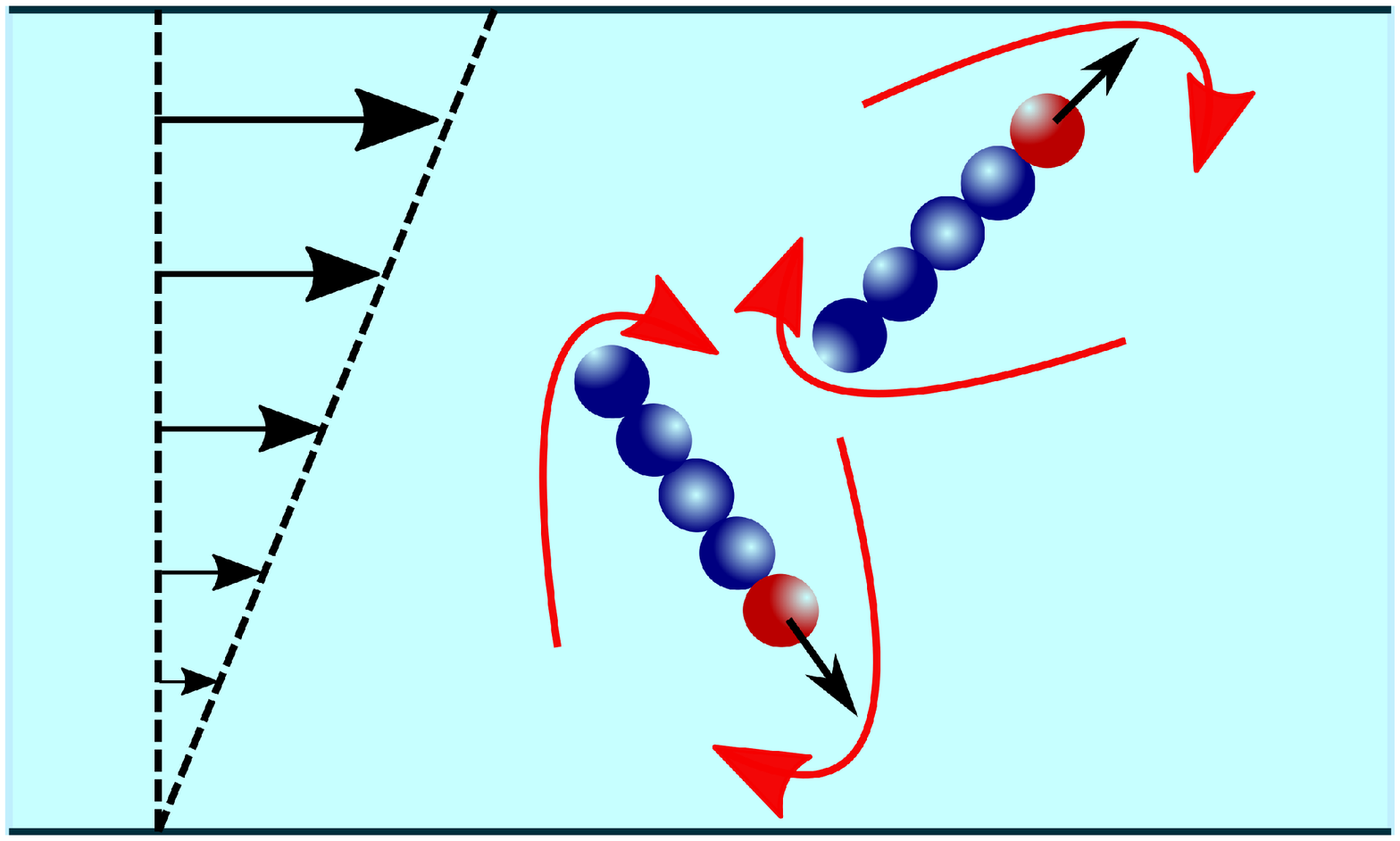}
  	\caption{a) The variation of $\phi$ near the top wall $y'=1$ for $Pe=0.5$ for various  $Wi=0,0.75,1.5,7.5,30,75$ and $150$. b) Schematics of flow induced alignment of filaments on the top and bottom half of the channel.}
  	\label{Fig:cosphi}
  \end{figure}
 
 Now, we turn our attention to the flow driven angular distribution of the filament in close vicinity of the walls ($\Delta y=l/2$). As expected, the angle distribution $P(\phi)$ near the top wall is symmetric about $\pi/2$, which consists two identical peaks centred at $\phi=0$, and $\phi=\pi$ at  $Wi=0$ (see Fig.\ref{Fig:cosphi}-a). Figure displays the distribution of $\phi$ at a fixed $Pe=0.5$ for several values of $Wi$. The peak height at $\phi=0$ (major) increases and $\phi=\pi$ (minor) decreases with $Wi$. The height of peak shows a non-monotonic behavior, which exhibits a decrease followed by an increase in the limit of large $Wi$ at $\pi$.
 
 Summarising above results here, the active filaments align themselves along (against) the flow direction at the top (bottom) wall assisted by the torque due to shear-force in the P{\'e}clet number dominant regime as shown in Fig.~\ref{Fig:cosphi}-b. However, in the shear-dominant regime, i.e., $Wi >>1$, filament also aligns against (along) the flow at the top (bottom) surface.
 
 %The shear-induced alignment is identifies a critical point from the onset of the decrease to increase of the peak height at $\phi=\pi$. 
  
\subsection{Population of filaments at surface}
\begin{figure}%[h]
 	\includegraphics[width=\linewidth]{topwall_ratio}
 	\includegraphics[width=1.1\linewidth]{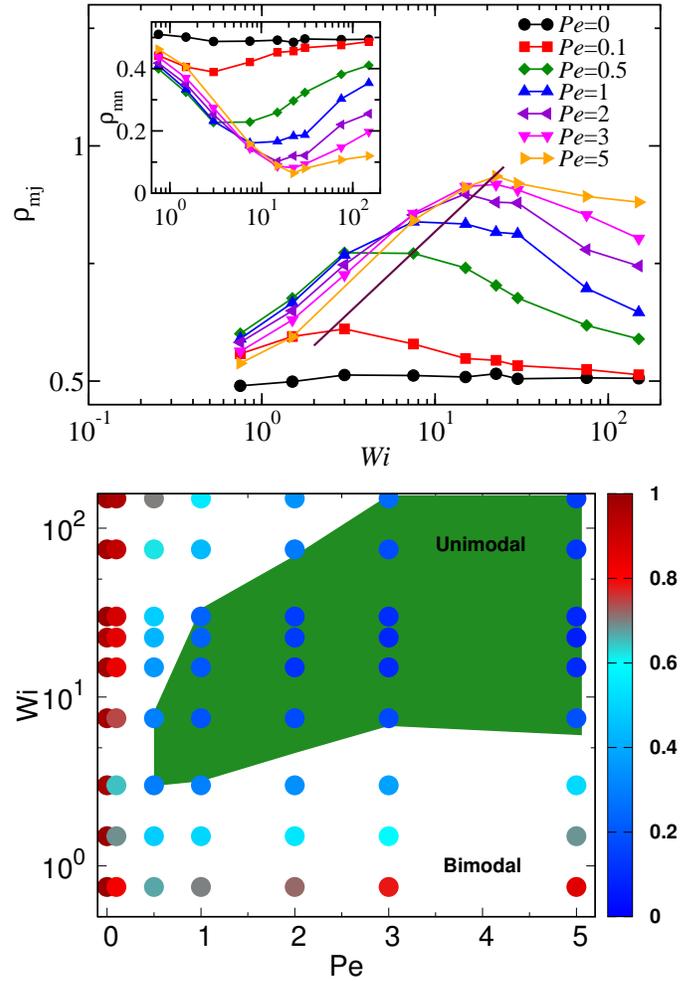}
 	\caption{a) Number density of majority population parallel to the flow at top wall. Inset shows the fraction of minority population of downstream to the total population rods at the top wall. A solid line shows the onset of the decrease in the majority population. 
 	b) The figure shows unimodal and bimodal phases in parameter space of $Pe$ and $Wi$.  The green shaded area shows the unimodal phase in the $Wi$ and $Pe$, similarly white area shows the bimodal phase.}
 	\label{near_wall}
 \end{figure}

 We address here population of upstream and downstream swimmers, especially near the surfaces. The coupling between  torque (due to flow) and  propulsive force leads  downstream to be the majority population and upstream to be the minority population near the top wall (see for Fig.\ref{Fig:cosphi}-b). Figure\ref{near_wall}-a displays the fraction of majority population ($\rho_{mj}= N_{mj}/N_p^w$) with respect to the total population $N_p^w$ at $y=L_y$, here $N_{mj}$ is number of filaments aligned along flow. The fraction $\rho_{mj}$   grows in small shear limit (all $Pe$) followed by a reduction in the limit of $Wi/Pe>6$, which continues to decrease with $Wi/Pe>>6$. The flow assists alignment, therefore the majority population increases in the limit of $Wi/Pe<6$. Further in the limit of large shear, tumbling motion leads to an increase in the effective rotational diffusion, as a result, the alignment against the flow  increases.  This results a decline in the majority population and increase in the minority population.
 
 The magnitude of shear force required to decrease the majority population for higher activity strength is substantially large, as we can see that the location of peak shifts towards higher $Wi$ with $Pe$ (see  Fig.~\ref{near_wall}-a). The minority fraction, $\rho_{mn}= N_{mn}/N_p^w$, at the top wall displays the same trend as shown in  inset of Fig.\ref{near_wall}-a. Here the decrease in population is  succeeded by an increase for all $Pe$. The behavior is identical on both surfaces. The difference in the preference of orientation at the top and bottom surfaces is due to clockwise torque on the filament, as Fig.\ref{Fig:cosphi}-b illustrates, which favours downstream on the top wall and upstream on the bottom wall. {\cblue The critical value of shear-rate, where the onset of increase in minority population at the wall occurs, is displayed from a solid line in Fig.~\ref{near_wall}.  This is known as onset of population splitting in the literature.}

{\cblue
Based on the above summary, we have identified a phase-diagram in $Pe$ and $Wi$ parameter space. This can be categorized in the three regimes. 
 i) Active force dominant regime,i.e., at small $Wi$, where swimming speed $Pe$ plays the sole role. Here the population of upstream and downstream are nearly same at the surfaces, thus we call it the biomodal phase. It is displayed below the green shaded area in Fig.~\ref{near_wall}-b.
 ii) Intermediate regime, where shear and active forces compete with each other, leading to  dominance of the majority population  over  the minority ( $Pe>0.5$ and $Wi/Pe \ge 6$). We define a unimodal phase, if majority population exceeds more than $80\%$ out of all population on the wall. The unimodal phase is displayed in green shaded region in Fig.~\ref{near_wall}-b. Note that there are still some minority population on the wall but they are negligible as  Fig.~\ref{Fig:orient:heat}-c and d illustrates. 
 iii) At sufficiently high Weissenberg number, the flow dominates over the  activity and the effect of self-propulsion is negligible. Hence, filaments behave like a passive one, and it leads to equal population of upstream and downstream filaments on the walls for $Wi/Pe>>6$. The transition from a unimodal to bimodal appears again as shown in Fig.~\ref{near_wall}-b above the green shaded region. If the minority population exceeds  $20\%$, we call it  bimodal phase.  The color bar in Fig.~\ref{near_wall} shows the ratio of the  minority to majority population.  In the diffusive regime $Pe<<1$, system is always in the bimodal phase. }

 %{\cblue Based on the above summary, we have identified a phase-diagram  in $Pe$ and $Wi$ parameter space. A system is said to be in the  unimodal phase if it share  shares more than $80\%$ to be majority population out of all near the wall. Here  majority population dominates. The  green shaded area in Fig.~\label{near_wall}-b   reflects the ratio of minority to majority population. The green shaded region in Fig.~\label{near_wall}-b unimodal population. As we move away from the shaded area, the minority population grows. 
% Although the fraction of majority population may be higher than the minority population, but their population has started declining with flow. This eventually gives a bimodal population of upstream and downstream filaments. Thus in shear dominant regime, swimmer do upstream and downstream swimming tending towards the equal weight throughout the channel.}
 
 \begin{figure}%[h]
 	\includegraphics*[width=\linewidth]{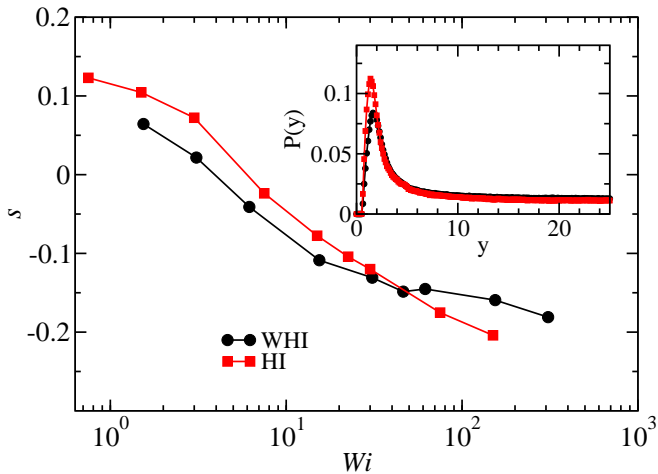}
 	\caption{Comparison of surface excess of the filament with (squares) and without (bullets) HI at $Pe=1$. Inset shows the distribution of rods as a function of distance from the wall at $Wi=0$ for  $Pe=1$.}
 	\label{Fig:HI-WHI}
 \end{figure} 
 
 \subsection{Effect of hydrodynamics}
  
 In this section, we compare the role of hydrodynamic correlations on the accumulation of confined active filaments. We have performed simulations with the random MPC, which has similar background solvent. In this method,  the long-range correlations among solvents are absent, however  it exhibits same transport properties as MPC fluids\cite{kikuchi2002polymer,kikuchi2003transport,ripoll2007hydrodynamic}. The comparison of surface adsorption with and without HI is demonstrated in Fig.~\ref{Fig:HI-WHI} for the same $Pe$, which suggests that the adsorption is enhanced with hydrodynamics in weak shear limit. The inset illustrates the distribution of swimmers as a function of distance from the wall, which clearly shows a substantial difference in accumulation with and without HI. The role of flow on surface excess is also shown in Fig.\ref{Fig:HI-WHI}, which is higher with hydrodynamics for small $Wi$. The enhancement in the adsorption with HI occurs due to increase in inhomogeneous drag force and interaction of solvents with wall\cite{padding2010translational,lin2000direct}. Another important aspect needs to be mentioned here that the difference in surface excess diminishes in the intermediate shear-rate. In this limit, transport due to transverse diffusive motion is dominant thus a swimmer moves across the channel relatively slower due to small transverse diffusivity with HI for smaller filaments, which gives smaller surface excess.  
  
 % {\cblue Furthermore, we also compare the population splitting of the  majority and minority swimmers. The qualitative behavior of both are same however quantitatively they are different. The shear dominance region appears nearly an order of magnitude larger value of $Wi$ without HI (see inset of Fig\ref{HI-WHI}). However fraction of majority swimmers near the flow wall is always larger relative to HI  for the situation of $Wi>1$. Hydrodynamic correlations assist for higher minority population, which is reflected also in  shift of onset of the peak towards higher $Wi$ without HI. }  
  
 \section{Summary and Conclusion}
 In summary, we have presented dynamics of a dilute concentration of active filaments confined between parallel walls under linear shear flow using hybrid MD-MPC simulations. Active filaments display a strong tendency to accumulate on the wall and shear  weakens this accumulation. The adsorption and desorption are analysed in terms of  alignment across the channel, anisotropic friction of the filament, diffusion across the channel, collision with wall, and residence time. The filaments  are aligned near the surfaces, and  anisotropic nature of friction\cite{elgeti2015run,doi1988theory} causes  slow short time translational and rotational diffusion across the channel. These mechanisms  result relatively larger residence time near the wall, hence higher adsorption without flow. Similarly, shear induced angular alignment \cite{Huang:Macromol2010,Singh_2014_JCP,singh2013dynamical} suppresses the ballistic motion across the channel leading to small concentration.
   
 The  surface excess "$s$" follows a power-law behavior as $s\sim Wi^{-\beta}$, with $\beta\approx .8$, in the intermediate regime of flow. The simulation results are also validated with the help of scaling arguments, which predicts nearly same exponent $\beta=7/9$. The adsorption is sufficiently suppressed by flow in the limit of $Wi>>1$, and density profile of active filament is similar to passive system with weak adsorption on the surfaces. The adsorption and desorption is also quantified with the help of residence time near the wall and bulk $\tau_w/\tau_b$, which exhibits similar dependence. The suppression of angular fluctuations causes slow decrease in the bulk residence time, thus the adsorption decreases with shear.
    
 The angular alignment of active filaments along the flow shows a non-monotonic distribution profile at the walls with $Wi$. The increase in angular alignment of upstream (downstream) swimmers at bottom (top) wall is followed by the onset of its decrease after a critical shear-rate at $\phi=0$ ($\phi=\pi$). In addition to that, the fraction of majority (upstream) population at the top wall also shows a non-monotonic behavior with $Wi$. The local angular distributions near the surfaces lead to a power law variation as $\chi_{p} \sim Wi^{-\delta}$, with an exponent $\delta\sim 1/5$. The smaller exponent than the bulk suggests weaker relative variation on the surfaces. More importantly, the orientational moment decreases with propulsive force near the wall unlike the bulk where it is independent of P{\'e}clet number. The dependence of $\chi_p$ is associated with the steep steric interactions with walls. The onset of decrease in the majority population illustrates the  weakening of  effect of self-propulsion over high shear forces, which eventually leads to equal population at angular separation $\pi$\cite{Nili:rsc:2017}.
 
 The dynamics of modelled microswimmers in complex active environments would be interesting to consider in the future studies, this may bring close to more realistic systems. More-specifically, curved and soft  surfaces, shear-gradients, chiral-shaped swimmers,  and viscoelastic media may lead to several distinguishable motional phases. {\cblue It would be further interesting to consider conservation of angular momentum in such simulations, especially for a chiral-shaped or spherically symmetric swimmers.  It may influence the stream lines near the swimmers, motility and density-profile of swimmers\cite{yang2015effect}.} 
 
 \section{Acknowledgements}
 Authors acknowledge HPC facility at IISER Bhopal for computation time. We thank DST SERB Grant No. 
 YSS/2015/000230 for the financial support. SKA thanks IISER Bhopal for the funding.

%\bibliographystyle{apsrev}
%\bibliography{references}

\begin{thebibliography}{95}
	\expandafter\ifx\csname natexlab\endcsname\relax\def\natexlab#1{#1}\fi
	\expandafter\ifx\csname bibnamefont\endcsname\relax
	\def\bibnamefont#1{#1}\fi
	\expandafter\ifx\csname bibfnamefont\endcsname\relax
	\def\bibfnamefont#1{#1}\fi
	\expandafter\ifx\csname citenamefont\endcsname\relax
	\def\citenamefont#1{#1}\fi
	\expandafter\ifx\csname url\endcsname\relax
	\def\url#1{\texttt{#1}}\fi
	\expandafter\ifx\csname urlprefix\endcsname\relax\def\urlprefix{URL }\fi
	\providecommand{\bibinfo}[2]{#2}
	\providecommand{\eprint}[2][]{\url{#2}}
	
	\bibitem[{\citenamefont{Ringo}(1967)}]{Ringo543}
	\bibinfo{author}{\bibfnamefont{D.~L.} \bibnamefont{Ringo}},
	\bibinfo{journal}{The Journal of Cell Biology} \textbf{\bibinfo{volume}{33}},
	\bibinfo{pages}{543} (\bibinfo{year}{1967}), ISSN \bibinfo{issn}{0021-9525}.
	
	\bibitem[{\citenamefont{Polin et~al.}(2009)\citenamefont{Polin, Tuval,
			Drescher, Gollub, and Goldstein}}]{Polin487}
	\bibinfo{author}{\bibfnamefont{M.}~\bibnamefont{Polin}},
	\bibinfo{author}{\bibfnamefont{I.}~\bibnamefont{Tuval}},
	\bibinfo{author}{\bibfnamefont{K.}~\bibnamefont{Drescher}},
	\bibinfo{author}{\bibfnamefont{J.~P.} \bibnamefont{Gollub}},
	\bibnamefont{and} \bibinfo{author}{\bibfnamefont{R.~E.}
		\bibnamefont{Goldstein}}, \bibinfo{journal}{Science}
	\textbf{\bibinfo{volume}{325}}, \bibinfo{pages}{487} (\bibinfo{year}{2009}),
	ISSN \bibinfo{issn}{0036-8075}.
	
	\bibitem[{\citenamefont{Blair}(1995)}]{Blair1995}
	\bibinfo{author}{\bibfnamefont{D.~F.} \bibnamefont{Blair}},
	\bibinfo{journal}{Annual Review of Microbiology}
	\textbf{\bibinfo{volume}{49}}, \bibinfo{pages}{489} (\bibinfo{year}{1995}),
	\bibinfo{note}{pMID: 8561469}.
	
	\bibitem[{\citenamefont{BERG}(1973)}]{Berg1973}
	\bibinfo{author}{\bibfnamefont{H.~C.} \bibnamefont{BERG}},
	\bibinfo{journal}{Nature} \textbf{\bibinfo{volume}{245}},
	\bibinfo{pages}{380} (\bibinfo{year}{1973}).
	
	\bibitem[{\citenamefont{GRAY}(1955)}]{GRAY775}
	\bibinfo{author}{\bibfnamefont{J.}~\bibnamefont{GRAY}},
	\bibinfo{journal}{Journal of Experimental Biology}
	\textbf{\bibinfo{volume}{32}}, \bibinfo{pages}{775} (\bibinfo{year}{1955}),
	ISSN \bibinfo{issn}{0022-0949}.
	
	\bibitem[{\citenamefont{Shack et~al.}(1974)\citenamefont{Shack, Fray, and
			Lardner}}]{SHACK1974555}
	\bibinfo{author}{\bibfnamefont{W.}~\bibnamefont{Shack}},
	\bibinfo{author}{\bibfnamefont{C.}~\bibnamefont{Fray}}, \bibnamefont{and}
	\bibinfo{author}{\bibfnamefont{T.}~\bibnamefont{Lardner}},
	\bibinfo{journal}{Bulletin of Mathematical Biology}
	\textbf{\bibinfo{volume}{36}}, \bibinfo{pages}{555 } (\bibinfo{year}{1974}),
	ISSN \bibinfo{issn}{0092-8240}.
	
	\bibitem[{\citenamefont{Woolley}(2003)}]{Woolley01082003}
	\bibinfo{author}{\bibfnamefont{D.}~\bibnamefont{Woolley}},
	\bibinfo{journal}{Reproduction} \textbf{\bibinfo{volume}{126}},
	\bibinfo{pages}{259} (\bibinfo{year}{2003}).
	
	\bibitem[{\citenamefont{Tung et~al.}(2014)\citenamefont{Tung, Ardon, Fiore,
			Suarez, and Wu}}]{Tung:softmatt2014}
	\bibinfo{author}{\bibfnamefont{C.-k.} \bibnamefont{Tung}},
	\bibinfo{author}{\bibfnamefont{F.}~\bibnamefont{Ardon}},
	\bibinfo{author}{\bibfnamefont{A.~G.} \bibnamefont{Fiore}},
	\bibinfo{author}{\bibfnamefont{S.~S.} \bibnamefont{Suarez}},
	\bibnamefont{and} \bibinfo{author}{\bibfnamefont{M.}~\bibnamefont{Wu}},
	\bibinfo{journal}{Lab Chip} \textbf{\bibinfo{volume}{14}},
	\bibinfo{pages}{1348} (\bibinfo{year}{2014}).
	
	\bibitem[{\citenamefont{Lauga and Powers}(2009)}]{lauga2009hydrodynamics}
	\bibinfo{author}{\bibfnamefont{E.}~\bibnamefont{Lauga}} \bibnamefont{and}
	\bibinfo{author}{\bibfnamefont{T.~R.} \bibnamefont{Powers}},
	\bibinfo{journal}{Reports on Progress in Physics}
	\textbf{\bibinfo{volume}{72}}, \bibinfo{pages}{096601}
	(\bibinfo{year}{2009}).
	
	\bibitem[{\citenamefont{Uspal et~al.}(2015)\citenamefont{Uspal, Popescu,
			Dietrich, and Tasinkevych}}]{uspal2015rheotaxis}
	\bibinfo{author}{\bibfnamefont{W.}~\bibnamefont{Uspal}},
	\bibinfo{author}{\bibfnamefont{M.~N.} \bibnamefont{Popescu}},
	\bibinfo{author}{\bibfnamefont{S.}~\bibnamefont{Dietrich}}, \bibnamefont{and}
	\bibinfo{author}{\bibfnamefont{M.}~\bibnamefont{Tasinkevych}},
	\bibinfo{journal}{Soft Matter} \textbf{\bibinfo{volume}{11}},
	\bibinfo{pages}{6613} (\bibinfo{year}{2015}).
	
	\bibitem[{\citenamefont{Kantsler et~al.}(2014)\citenamefont{Kantsler, Dunkel,
			Blayney, and Goldstein}}]{kantsler2014rheotaxis}
	\bibinfo{author}{\bibfnamefont{V.}~\bibnamefont{Kantsler}},
	\bibinfo{author}{\bibfnamefont{J.}~\bibnamefont{Dunkel}},
	\bibinfo{author}{\bibfnamefont{M.}~\bibnamefont{Blayney}}, \bibnamefont{and}
	\bibinfo{author}{\bibfnamefont{R.~E.} \bibnamefont{Goldstein}},
	\bibinfo{journal}{Elife} \textbf{\bibinfo{volume}{3}},
	\bibinfo{pages}{e02403} (\bibinfo{year}{2014}).
	
	\bibitem[{\citenamefont{Rosengarten et~al.}(1988)\citenamefont{Rosengarten,
			Klein-Struckmeier, and Kirchhoff}}]{rosengarten1988rheotactic}
	\bibinfo{author}{\bibfnamefont{R.}~\bibnamefont{Rosengarten}},
	\bibinfo{author}{\bibfnamefont{A.}~\bibnamefont{Klein-Struckmeier}},
	\bibnamefont{and}
	\bibinfo{author}{\bibfnamefont{H.}~\bibnamefont{Kirchhoff}},
	\bibinfo{journal}{Journal of bacteriology} \textbf{\bibinfo{volume}{170}},
	\bibinfo{pages}{989} (\bibinfo{year}{1988}).
	
	\bibitem[{\citenamefont{Gao et~al.}(2012)\citenamefont{Gao, Kagan, Pak,
			Clawson, Campuzano, Chuluun-Erdene, Shipton, Fullerton, Zhang, Lauga
			et~al.}}]{gao2012cargo}
	\bibinfo{author}{\bibfnamefont{W.}~\bibnamefont{Gao}},
	\bibinfo{author}{\bibfnamefont{D.}~\bibnamefont{Kagan}},
	\bibinfo{author}{\bibfnamefont{O.~S.} \bibnamefont{Pak}},
	\bibinfo{author}{\bibfnamefont{C.}~\bibnamefont{Clawson}},
	\bibinfo{author}{\bibfnamefont{S.}~\bibnamefont{Campuzano}},
	\bibinfo{author}{\bibfnamefont{E.}~\bibnamefont{Chuluun-Erdene}},
	\bibinfo{author}{\bibfnamefont{E.}~\bibnamefont{Shipton}},
	\bibinfo{author}{\bibfnamefont{E.~E.} \bibnamefont{Fullerton}},
	\bibinfo{author}{\bibfnamefont{L.}~\bibnamefont{Zhang}},
	\bibinfo{author}{\bibfnamefont{E.}~\bibnamefont{Lauga}},
	\bibnamefont{et~al.}, \bibinfo{journal}{small} \textbf{\bibinfo{volume}{8}},
	\bibinfo{pages}{460} (\bibinfo{year}{2012}).
	
	\bibitem[{\citenamefont{Berke et~al.}(2008{\natexlab{a}})\citenamefont{Berke,
			Turner, Berg, and Lauga}}]{Berke:prl:2008}
	\bibinfo{author}{\bibfnamefont{A.~P.} \bibnamefont{Berke}},
	\bibinfo{author}{\bibfnamefont{L.}~\bibnamefont{Turner}},
	\bibinfo{author}{\bibfnamefont{H.~C.} \bibnamefont{Berg}}, \bibnamefont{and}
	\bibinfo{author}{\bibfnamefont{E.}~\bibnamefont{Lauga}},
	\bibinfo{journal}{Phys. Rev. Lett.} \textbf{\bibinfo{volume}{101}},
	\bibinfo{pages}{038102} (\bibinfo{year}{2008}{\natexlab{a}}).
	
	\bibitem[{\citenamefont{Sabass and Seifert}(2010)}]{sabass:prl:2010}
	\bibinfo{author}{\bibfnamefont{B.}~\bibnamefont{Sabass}} \bibnamefont{and}
	\bibinfo{author}{\bibfnamefont{U.}~\bibnamefont{Seifert}},
	\bibinfo{journal}{Phys. Rev. Lett.} \textbf{\bibinfo{volume}{105}},
	\bibinfo{pages}{218103} (\bibinfo{year}{2010}).
	
	\bibitem[{\citenamefont{Schaar et~al.}(2015)\citenamefont{Schaar, Z\"ottl, and
			Stark}}]{Schaar:PRL:2015}
	\bibinfo{author}{\bibfnamefont{K.}~\bibnamefont{Schaar}},
	\bibinfo{author}{\bibfnamefont{A.}~\bibnamefont{Z\"ottl}}, \bibnamefont{and}
	\bibinfo{author}{\bibfnamefont{H.}~\bibnamefont{Stark}},
	\bibinfo{journal}{Phys. Rev. Lett.} \textbf{\bibinfo{volume}{115}},
	\bibinfo{pages}{038101} (\bibinfo{year}{2015}).
	
	\bibitem[{\citenamefont{Elgeti and Gompper}(2015)}]{elgeti2015run}
	\bibinfo{author}{\bibfnamefont{J.}~\bibnamefont{Elgeti}} \bibnamefont{and}
	\bibinfo{author}{\bibfnamefont{G.}~\bibnamefont{Gompper}},
	\bibinfo{journal}{EPL (Europhysics Letters)} \textbf{\bibinfo{volume}{109}},
	\bibinfo{pages}{58003} (\bibinfo{year}{2015}).
	
	\bibitem[{\citenamefont{Tournus et~al.}(2015)\citenamefont{Tournus, Kirshtein,
			Berlyand, and Aranson}}]{tournus2015flexibility}
	\bibinfo{author}{\bibfnamefont{M.}~\bibnamefont{Tournus}},
	\bibinfo{author}{\bibfnamefont{A.}~\bibnamefont{Kirshtein}},
	\bibinfo{author}{\bibfnamefont{L.}~\bibnamefont{Berlyand}}, \bibnamefont{and}
	\bibinfo{author}{\bibfnamefont{I.~S.} \bibnamefont{Aranson}},
	\bibinfo{journal}{Journal of the Royal Society Interface}
	\textbf{\bibinfo{volume}{12}}, \bibinfo{pages}{20140904}
	(\bibinfo{year}{2015}).
	
	\bibitem[{\citenamefont{Das et~al.}(2018)\citenamefont{Das, Gompper, and
			Winkler}}]{das2018confined}
	\bibinfo{author}{\bibfnamefont{S.}~\bibnamefont{Das}},
	\bibinfo{author}{\bibfnamefont{G.}~\bibnamefont{Gompper}}, \bibnamefont{and}
	\bibinfo{author}{\bibfnamefont{R.~G.} \bibnamefont{Winkler}},
	\bibinfo{journal}{New Journal of Physics} \textbf{\bibinfo{volume}{20}},
	\bibinfo{pages}{015001} (\bibinfo{year}{2018}).
	
	\bibitem[{\citenamefont{Daddi-Moussa-Ider
			et~al.}(2018)\citenamefont{Daddi-Moussa-Ider, Lisicki, Hoell, and
			L{\"o}wen}}]{daddi2018swimming}
	\bibinfo{author}{\bibfnamefont{A.}~\bibnamefont{Daddi-Moussa-Ider}},
	\bibinfo{author}{\bibfnamefont{M.}~\bibnamefont{Lisicki}},
	\bibinfo{author}{\bibfnamefont{C.}~\bibnamefont{Hoell}}, \bibnamefont{and}
	\bibinfo{author}{\bibfnamefont{H.}~\bibnamefont{L{\"o}wen}},
	\bibinfo{journal}{The Journal of chemical physics}
	\textbf{\bibinfo{volume}{148}}, \bibinfo{pages}{134904}
	(\bibinfo{year}{2018}).
	
	\bibitem[{\citenamefont{Ledesma-Aguilar
			et~al.}(2012)\citenamefont{Ledesma-Aguilar, L{\"o}wen, and
			Yeomans}}]{ledesma2012circle}
	\bibinfo{author}{\bibfnamefont{R.}~\bibnamefont{Ledesma-Aguilar}},
	\bibinfo{author}{\bibfnamefont{H.}~\bibnamefont{L{\"o}wen}},
	\bibnamefont{and} \bibinfo{author}{\bibfnamefont{J.~M.}
		\bibnamefont{Yeomans}}, \bibinfo{journal}{The European Physical Journal E}
	\textbf{\bibinfo{volume}{35}}, \bibinfo{pages}{70} (\bibinfo{year}{2012}).
	
	\bibitem[{\citenamefont{Elgeti and Gompper}(2016)}]{Elgeti2016}
	\bibinfo{author}{\bibfnamefont{J.}~\bibnamefont{Elgeti}} \bibnamefont{and}
	\bibinfo{author}{\bibfnamefont{G.}~\bibnamefont{Gompper}},
	\bibinfo{journal}{The European Physical Journal Special Topics}
	\textbf{\bibinfo{volume}{225}}, \bibinfo{pages}{2333} (\bibinfo{year}{2016}),
	ISSN \bibinfo{issn}{1951-6401}.
	
	\bibitem[{\citenamefont{Potomkin et~al.}(2017)\citenamefont{Potomkin, Kaiser,
			Berlyand, and Aranson}}]{potomkin2017focusing}
	\bibinfo{author}{\bibfnamefont{M.}~\bibnamefont{Potomkin}},
	\bibinfo{author}{\bibfnamefont{A.}~\bibnamefont{Kaiser}},
	\bibinfo{author}{\bibfnamefont{L.}~\bibnamefont{Berlyand}}, \bibnamefont{and}
	\bibinfo{author}{\bibfnamefont{I.}~\bibnamefont{Aranson}},
	\bibinfo{journal}{New Journal of Physics} \textbf{\bibinfo{volume}{19}},
	\bibinfo{pages}{115005} (\bibinfo{year}{2017}).
	
	\bibitem[{\citenamefont{Omori and Ishikawa}(2016)}]{omori2016upward}
	\bibinfo{author}{\bibfnamefont{T.}~\bibnamefont{Omori}} \bibnamefont{and}
	\bibinfo{author}{\bibfnamefont{T.}~\bibnamefont{Ishikawa}},
	\bibinfo{journal}{Physical Review E} \textbf{\bibinfo{volume}{93}},
	\bibinfo{pages}{032402} (\bibinfo{year}{2016}).
	
	\bibitem[{\citenamefont{Tao and Kapral}(2010)}]{tao2010swimming}
	\bibinfo{author}{\bibfnamefont{Y.-G.} \bibnamefont{Tao}} \bibnamefont{and}
	\bibinfo{author}{\bibfnamefont{R.}~\bibnamefont{Kapral}},
	\bibinfo{journal}{Soft Matter} \textbf{\bibinfo{volume}{6}},
	\bibinfo{pages}{756} (\bibinfo{year}{2010}).
	
	\bibitem[{\citenamefont{de~Graaf et~al.}(2016)\citenamefont{de~Graaf,
			Mathijssen, Fabritius, Menke, Holm, and Shendruk}}]{de2016understanding}
	\bibinfo{author}{\bibfnamefont{J.}~\bibnamefont{de~Graaf}},
	\bibinfo{author}{\bibfnamefont{A.~J.} \bibnamefont{Mathijssen}},
	\bibinfo{author}{\bibfnamefont{M.}~\bibnamefont{Fabritius}},
	\bibinfo{author}{\bibfnamefont{H.}~\bibnamefont{Menke}},
	\bibinfo{author}{\bibfnamefont{C.}~\bibnamefont{Holm}}, \bibnamefont{and}
	\bibinfo{author}{\bibfnamefont{T.~N.} \bibnamefont{Shendruk}},
	\bibinfo{journal}{Soft Matter} \textbf{\bibinfo{volume}{12}},
	\bibinfo{pages}{4704} (\bibinfo{year}{2016}).
	
	\bibitem[{\citenamefont{Pagonabarraga and
			Llopis}(2013)}]{pagonabarraga2013structure}
	\bibinfo{author}{\bibfnamefont{I.}~\bibnamefont{Pagonabarraga}}
	\bibnamefont{and} \bibinfo{author}{\bibfnamefont{I.}~\bibnamefont{Llopis}},
	\bibinfo{journal}{Soft Matter} \textbf{\bibinfo{volume}{9}},
	\bibinfo{pages}{7174} (\bibinfo{year}{2013}).
	
	\bibitem[{\citenamefont{Zhang et~al.}(2010)\citenamefont{Zhang, Petit, Lu,
			Kratochvil, Peyer, Pei, Lou, and Nelson}}]{zhang:acs:2010}
	\bibinfo{author}{\bibfnamefont{L.}~\bibnamefont{Zhang}},
	\bibinfo{author}{\bibfnamefont{T.}~\bibnamefont{Petit}},
	\bibinfo{author}{\bibfnamefont{Y.}~\bibnamefont{Lu}},
	\bibinfo{author}{\bibfnamefont{B.~E.} \bibnamefont{Kratochvil}},
	\bibinfo{author}{\bibfnamefont{K.~E.} \bibnamefont{Peyer}},
	\bibinfo{author}{\bibfnamefont{R.}~\bibnamefont{Pei}},
	\bibinfo{author}{\bibfnamefont{J.}~\bibnamefont{Lou}}, \bibnamefont{and}
	\bibinfo{author}{\bibfnamefont{B.~J.} \bibnamefont{Nelson}},
	\bibinfo{journal}{ACS Nano} \textbf{\bibinfo{volume}{4}},
	\bibinfo{pages}{6228} (\bibinfo{year}{2010}), \bibinfo{note}{pMID: 20873764}.
	
	\bibitem[{\citenamefont{Hill et~al.}(2007)\citenamefont{Hill, Kalkanci,
			McMurry, and Koser}}]{Hill:PRL2007}
	\bibinfo{author}{\bibfnamefont{J.}~\bibnamefont{Hill}},
	\bibinfo{author}{\bibfnamefont{O.}~\bibnamefont{Kalkanci}},
	\bibinfo{author}{\bibfnamefont{J.~L.} \bibnamefont{McMurry}},
	\bibnamefont{and} \bibinfo{author}{\bibfnamefont{H.}~\bibnamefont{Koser}},
	\bibinfo{journal}{Phys. Rev. Lett.} \textbf{\bibinfo{volume}{98}},
	\bibinfo{pages}{068101} (\bibinfo{year}{2007}).
	
	\bibitem[{\citenamefont{Kaya and Koser}(2012)}]{KAYABiophysical:2012}
	\bibinfo{author}{\bibfnamefont{T.}~\bibnamefont{Kaya}} \bibnamefont{and}
	\bibinfo{author}{\bibfnamefont{H.}~\bibnamefont{Koser}},
	\bibinfo{journal}{Biophysical Journal} \textbf{\bibinfo{volume}{102}},
	\bibinfo{pages}{1514 } (\bibinfo{year}{2012}), ISSN
	\bibinfo{issn}{0006-3495}.
	
	\bibitem[{\citenamefont{Yuan et~al.}(2015)\citenamefont{Yuan, Raizen, and
			Bau}}]{Yuan:PNAS2015}
	\bibinfo{author}{\bibfnamefont{J.}~\bibnamefont{Yuan}},
	\bibinfo{author}{\bibfnamefont{D.~M.} \bibnamefont{Raizen}},
	\bibnamefont{and} \bibinfo{author}{\bibfnamefont{H.~H.} \bibnamefont{Bau}},
	\bibinfo{journal}{Proceedings of the National Academy of Sciences}
	\textbf{\bibinfo{volume}{112}}, \bibinfo{pages}{3606} (\bibinfo{year}{2015}),
	ISSN \bibinfo{issn}{0027-8424}.
	
	\bibitem[{\citenamefont{Howse et~al.}(2007)\citenamefont{Howse, Jones, Ryan,
			Gough, Vafabakhsh, and Golestanian}}]{Howseprl2007}
	\bibinfo{author}{\bibfnamefont{J.~R.} \bibnamefont{Howse}},
	\bibinfo{author}{\bibfnamefont{R.~A.~L.} \bibnamefont{Jones}},
	\bibinfo{author}{\bibfnamefont{A.~J.} \bibnamefont{Ryan}},
	\bibinfo{author}{\bibfnamefont{T.}~\bibnamefont{Gough}},
	\bibinfo{author}{\bibfnamefont{R.}~\bibnamefont{Vafabakhsh}},
	\bibnamefont{and}
	\bibinfo{author}{\bibfnamefont{R.}~\bibnamefont{Golestanian}},
	\bibinfo{journal}{Phys. Rev. Lett.} \textbf{\bibinfo{volume}{99}},
	\bibinfo{pages}{048102} (\bibinfo{year}{2007}).
	
	\bibitem[{\citenamefont{Paxton et~al.}(2006)\citenamefont{Paxton, Sundararajan,
			Mallouk, and Sen}}]{Paxton2006}
	\bibinfo{author}{\bibfnamefont{W.~F.} \bibnamefont{Paxton}},
	\bibinfo{author}{\bibfnamefont{S.}~\bibnamefont{Sundararajan}},
	\bibinfo{author}{\bibfnamefont{T.~E.} \bibnamefont{Mallouk}},
	\bibnamefont{and} \bibinfo{author}{\bibfnamefont{A.}~\bibnamefont{Sen}},
	\bibinfo{journal}{Angewandte Chemie International Edition}
	\textbf{\bibinfo{volume}{45}}, \bibinfo{pages}{5420} (\bibinfo{year}{2006}),
	ISSN \bibinfo{issn}{1521-3773}.
	
	\bibitem[{\citenamefont{Palacci et~al.}(2015)\citenamefont{Palacci, Sacanna,
			Abramian, Barral, Hanson, Grosberg, Pine, and
			Chaikin}}]{palacci2015artificial}
	\bibinfo{author}{\bibfnamefont{J.}~\bibnamefont{Palacci}},
	\bibinfo{author}{\bibfnamefont{S.}~\bibnamefont{Sacanna}},
	\bibinfo{author}{\bibfnamefont{A.}~\bibnamefont{Abramian}},
	\bibinfo{author}{\bibfnamefont{J.}~\bibnamefont{Barral}},
	\bibinfo{author}{\bibfnamefont{K.}~\bibnamefont{Hanson}},
	\bibinfo{author}{\bibfnamefont{A.~Y.} \bibnamefont{Grosberg}},
	\bibinfo{author}{\bibfnamefont{D.~J.} \bibnamefont{Pine}}, \bibnamefont{and}
	\bibinfo{author}{\bibfnamefont{P.~M.} \bibnamefont{Chaikin}},
	\bibinfo{journal}{Science advances} \textbf{\bibinfo{volume}{1}},
	\bibinfo{pages}{e1400214} (\bibinfo{year}{2015}).
	
	\bibitem[{\citenamefont{Nili et~al.}(2017)\citenamefont{Nili, Kheyri, Abazari,
			Fahimniya, and Naji}}]{Nili:rsc:2017}
	\bibinfo{author}{\bibfnamefont{H.}~\bibnamefont{Nili}},
	\bibinfo{author}{\bibfnamefont{M.}~\bibnamefont{Kheyri}},
	\bibinfo{author}{\bibfnamefont{J.}~\bibnamefont{Abazari}},
	\bibinfo{author}{\bibfnamefont{A.}~\bibnamefont{Fahimniya}},
	\bibnamefont{and} \bibinfo{author}{\bibfnamefont{A.}~\bibnamefont{Naji}},
	\bibinfo{journal}{Soft Matter} \textbf{\bibinfo{volume}{13}},
	\bibinfo{pages}{4494} (\bibinfo{year}{2017}).
	
	\bibitem[{\citenamefont{Chilukuri et~al.}(2014)\citenamefont{Chilukuri,
			Collins, and Underhill}}]{Chilukuri:jpcm:2014}
	\bibinfo{author}{\bibfnamefont{S.}~\bibnamefont{Chilukuri}},
	\bibinfo{author}{\bibfnamefont{C.~H.} \bibnamefont{Collins}},
	\bibnamefont{and} \bibinfo{author}{\bibfnamefont{P.~T.}
		\bibnamefont{Underhill}}, \bibinfo{journal}{Journal of Physics: Condensed
		Matter} \textbf{\bibinfo{volume}{26}}, \bibinfo{pages}{115101}
	(\bibinfo{year}{2014}).
	
	\bibitem[{\citenamefont{Ezhilan and
			Saintillan}(2015)}]{ezhilan_saintillan_2015}
	\bibinfo{author}{\bibfnamefont{B.}~\bibnamefont{Ezhilan}} \bibnamefont{and}
	\bibinfo{author}{\bibfnamefont{D.}~\bibnamefont{Saintillan}},
	\bibinfo{journal}{Journal of Fluid Mechanics} \textbf{\bibinfo{volume}{777}},
	\bibinfo{pages}{482–522} (\bibinfo{year}{2015}).
	
	\bibitem[{\citenamefont{Bretherton and Rothschild}(1961)}]{Bretherton490}
	\bibinfo{author}{\bibnamefont{Bretherton}} \bibnamefont{and}
	\bibinfo{author}{\bibnamefont{Rothschild}}, \bibinfo{journal}{Proceedings of
		the Royal Society of London B: Biological Sciences}
	\textbf{\bibinfo{volume}{153}}, \bibinfo{pages}{490} (\bibinfo{year}{1961}),
	ISSN \bibinfo{issn}{0080-4649}.
	
	\bibitem[{\citenamefont{Zhang et~al.}(2016)\citenamefont{Zhang, Liu, Meriano,
			Ru, Xie, Luo, and Sun}}]{Zhang:srep:2016}
	\bibinfo{author}{\bibfnamefont{Z.}~\bibnamefont{Zhang}},
	\bibinfo{author}{\bibfnamefont{J.}~\bibnamefont{Liu}},
	\bibinfo{author}{\bibfnamefont{J.}~\bibnamefont{Meriano}},
	\bibinfo{author}{\bibfnamefont{C.}~\bibnamefont{Ru}},
	\bibinfo{author}{\bibfnamefont{S.}~\bibnamefont{Xie}},
	\bibinfo{author}{\bibfnamefont{J.}~\bibnamefont{Luo}}, \bibnamefont{and}
	\bibinfo{author}{\bibfnamefont{Y.}~\bibnamefont{Sun}},
	\bibinfo{journal}{Scientific Reports} \textbf{\bibinfo{volume}{6}},
	\bibinfo{pages}{23553 EP } (\bibinfo{year}{2016}), \bibinfo{note}{article}.
	
	\bibitem[{\citenamefont{Katuri et~al.}(2018)\citenamefont{Katuri, Uspal,
			Simmchen, Miguel-L{\'o}pez, and S{\'a}nchez}}]{katuri2018cross}
	\bibinfo{author}{\bibfnamefont{J.}~\bibnamefont{Katuri}},
	\bibinfo{author}{\bibfnamefont{W.~E.} \bibnamefont{Uspal}},
	\bibinfo{author}{\bibfnamefont{J.}~\bibnamefont{Simmchen}},
	\bibinfo{author}{\bibfnamefont{A.}~\bibnamefont{Miguel-L{\'o}pez}},
	\bibnamefont{and}
	\bibinfo{author}{\bibfnamefont{S.}~\bibnamefont{S{\'a}nchez}},
	\bibinfo{journal}{Science advances} \textbf{\bibinfo{volume}{4}},
	\bibinfo{pages}{eaao1755} (\bibinfo{year}{2018}).
	
	\bibitem[{\citenamefont{Son et~al.}(2015)\citenamefont{Son, Brumley, and
			Stocker}}]{son2015live}
	\bibinfo{author}{\bibfnamefont{K.}~\bibnamefont{Son}},
	\bibinfo{author}{\bibfnamefont{D.~R.} \bibnamefont{Brumley}},
	\bibnamefont{and} \bibinfo{author}{\bibfnamefont{R.}~\bibnamefont{Stocker}},
	\bibinfo{journal}{Nature Reviews Microbiology} \textbf{\bibinfo{volume}{13}},
	\bibinfo{pages}{761} (\bibinfo{year}{2015}).
	
	\bibitem[{\citenamefont{Rusconi et~al.}(2014)\citenamefont{Rusconi, Guasto, and
			Stocker}}]{rusconi2014bacterial}
	\bibinfo{author}{\bibfnamefont{R.}~\bibnamefont{Rusconi}},
	\bibinfo{author}{\bibfnamefont{J.~S.} \bibnamefont{Guasto}},
	\bibnamefont{and} \bibinfo{author}{\bibfnamefont{R.}~\bibnamefont{Stocker}},
	\bibinfo{journal}{Nature physics} \textbf{\bibinfo{volume}{10}},
	\bibinfo{pages}{212} (\bibinfo{year}{2014}).
	
	\bibitem[{\citenamefont{Kaya and Koser}(2009)}]{kaya2009characterization}
	\bibinfo{author}{\bibfnamefont{T.}~\bibnamefont{Kaya}} \bibnamefont{and}
	\bibinfo{author}{\bibfnamefont{H.}~\bibnamefont{Koser}},
	\bibinfo{journal}{Physical review letters} \textbf{\bibinfo{volume}{103}},
	\bibinfo{pages}{138103} (\bibinfo{year}{2009}).
	
	\bibitem[{\citenamefont{Meng et~al.}(2005)\citenamefont{Meng, Li, Galvani, Hao,
			Turner, Burr, and Hoch}}]{meng2005upstream}
	\bibinfo{author}{\bibfnamefont{Y.}~\bibnamefont{Meng}},
	\bibinfo{author}{\bibfnamefont{Y.}~\bibnamefont{Li}},
	\bibinfo{author}{\bibfnamefont{C.~D.} \bibnamefont{Galvani}},
	\bibinfo{author}{\bibfnamefont{G.}~\bibnamefont{Hao}},
	\bibinfo{author}{\bibfnamefont{J.~N.} \bibnamefont{Turner}},
	\bibinfo{author}{\bibfnamefont{T.~J.} \bibnamefont{Burr}}, \bibnamefont{and}
	\bibinfo{author}{\bibfnamefont{H.}~\bibnamefont{Hoch}},
	\bibinfo{journal}{Journal of Bacteriology} \textbf{\bibinfo{volume}{187}},
	\bibinfo{pages}{5560} (\bibinfo{year}{2005}).
	
	\bibitem[{\citenamefont{Montgomery et~al.}(1997)\citenamefont{Montgomery,
			Baker, and Carton}}]{Montgomery:nat:1997}
	\bibinfo{author}{\bibfnamefont{J.~C.} \bibnamefont{Montgomery}},
	\bibinfo{author}{\bibfnamefont{C.~F.} \bibnamefont{Baker}}, \bibnamefont{and}
	\bibinfo{author}{\bibfnamefont{A.~G.} \bibnamefont{Carton}},
	\bibinfo{journal}{Nature} \textbf{\bibinfo{volume}{389}}, \bibinfo{pages}{960
		EP } (\bibinfo{year}{1997}).
	
	\bibitem[{\citenamefont{P.}(1974)}]{ARNOLD:1974}
	\bibinfo{author}{\bibfnamefont{A.~G.} \bibnamefont{P.}},
	\bibinfo{journal}{Biological Reviews} \textbf{\bibinfo{volume}{49}},
	\bibinfo{pages}{515} (\bibinfo{year}{1974}),
	\eprint{https://onlinelibrary.wiley.com/doi/pdf/10.1111/j.1469-185X.1974.tb01173.x}.
	
	\bibitem[{\citenamefont{Marcos et~al.}(2012)\citenamefont{Marcos, Fu, Powers,
			and Stocker}}]{Marcos:NAS2012}
	\bibinfo{author}{\bibnamefont{Marcos}}, \bibinfo{author}{\bibfnamefont{H.~C.}
		\bibnamefont{Fu}}, \bibinfo{author}{\bibfnamefont{T.~R.}
		\bibnamefont{Powers}}, \bibnamefont{and}
	\bibinfo{author}{\bibfnamefont{R.}~\bibnamefont{Stocker}},
	\bibinfo{journal}{Proceedings of the National Academy of Sciences}
	\textbf{\bibinfo{volume}{109}}, \bibinfo{pages}{4780} (\bibinfo{year}{2012}),
	ISSN \bibinfo{issn}{0027-8424}.
	
	\bibitem[{\citenamefont{Li and Tang}(2009{\natexlab{a}})}]{Guanglai:PRL2009}
	\bibinfo{author}{\bibfnamefont{G.}~\bibnamefont{Li}} \bibnamefont{and}
	\bibinfo{author}{\bibfnamefont{J.~X.} \bibnamefont{Tang}},
	\bibinfo{journal}{Phys. Rev. Lett.} \textbf{\bibinfo{volume}{103}},
	\bibinfo{pages}{078101} (\bibinfo{year}{2009}{\natexlab{a}}).
	
	\bibitem[{\citenamefont{Lin et~al.}(2000)\citenamefont{Lin, Yu, and
			Rice}}]{lin2000direct}
	\bibinfo{author}{\bibfnamefont{B.}~\bibnamefont{Lin}},
	\bibinfo{author}{\bibfnamefont{J.}~\bibnamefont{Yu}}, \bibnamefont{and}
	\bibinfo{author}{\bibfnamefont{S.~A.} \bibnamefont{Rice}},
	\bibinfo{journal}{Physical Review E} \textbf{\bibinfo{volume}{62}},
	\bibinfo{pages}{3909} (\bibinfo{year}{2000}).
	
	\bibitem[{\citenamefont{Najafi and Golestanian}(2004)}]{najafi2004simple}
	\bibinfo{author}{\bibfnamefont{A.}~\bibnamefont{Najafi}} \bibnamefont{and}
	\bibinfo{author}{\bibfnamefont{R.}~\bibnamefont{Golestanian}},
	\bibinfo{journal}{Physical Review E} \textbf{\bibinfo{volume}{69}},
	\bibinfo{pages}{062901} (\bibinfo{year}{2004}).
	
	\bibitem[{\citenamefont{Pande and Smith}(2015)}]{pande2015forces}
	\bibinfo{author}{\bibfnamefont{J.}~\bibnamefont{Pande}} \bibnamefont{and}
	\bibinfo{author}{\bibfnamefont{A.-S.} \bibnamefont{Smith}},
	\bibinfo{journal}{Soft Matter} \textbf{\bibinfo{volume}{11}},
	\bibinfo{pages}{2364} (\bibinfo{year}{2015}).
	
	\bibitem[{\citenamefont{Babel et~al.}(2016)\citenamefont{Babel, L{\"o}wen, and
			Menzel}}]{babel2016dynamics}
	\bibinfo{author}{\bibfnamefont{S.}~\bibnamefont{Babel}},
	\bibinfo{author}{\bibfnamefont{H.}~\bibnamefont{L{\"o}wen}},
	\bibnamefont{and} \bibinfo{author}{\bibfnamefont{A.~M.}
		\bibnamefont{Menzel}}, \bibinfo{journal}{EPL (Europhysics Letters)}
	\textbf{\bibinfo{volume}{113}}, \bibinfo{pages}{58003}
	(\bibinfo{year}{2016}).
	
	\bibitem[{\citenamefont{Elgeti and Gompper}(2013{\natexlab{a}})}]{Elgeti:2013}
	\bibinfo{author}{\bibfnamefont{J.}~\bibnamefont{Elgeti}} \bibnamefont{and}
	\bibinfo{author}{\bibfnamefont{G.}~\bibnamefont{Gompper}},
	\bibinfo{journal}{EPL (Europhysics Letters)} \textbf{\bibinfo{volume}{101}},
	\bibinfo{pages}{48003} (\bibinfo{year}{2013}{\natexlab{a}}).
	
	\bibitem[{\citenamefont{{Elgeti, J.} and {Gompper, G.}}(2009)}]{Elgeti:2009}
	\bibinfo{author}{\bibnamefont{{Elgeti, J.}}} \bibnamefont{and}
	\bibinfo{author}{\bibnamefont{{Gompper, G.}}}, \bibinfo{journal}{EPL}
	\textbf{\bibinfo{volume}{85}}, \bibinfo{pages}{38002} (\bibinfo{year}{2009}).
	
	\bibitem[{\citenamefont{Tung et~al.}(2015)\citenamefont{Tung, Ardon, Roy, Koch,
			Suarez, and Wu}}]{Tung:PRL2015}
	\bibinfo{author}{\bibfnamefont{C.-k.} \bibnamefont{Tung}},
	\bibinfo{author}{\bibfnamefont{F.}~\bibnamefont{Ardon}},
	\bibinfo{author}{\bibfnamefont{A.}~\bibnamefont{Roy}},
	\bibinfo{author}{\bibfnamefont{D.~L.} \bibnamefont{Koch}},
	\bibinfo{author}{\bibfnamefont{S.~S.} \bibnamefont{Suarez}},
	\bibnamefont{and} \bibinfo{author}{\bibfnamefont{M.}~\bibnamefont{Wu}},
	\bibinfo{journal}{Phys. Rev. Lett.} \textbf{\bibinfo{volume}{114}},
	\bibinfo{pages}{108102} (\bibinfo{year}{2015}).
	
	\bibitem[{\citenamefont{Mart{\'\i}n-G{\'o}mez
			et~al.}(2018)\citenamefont{Mart{\'\i}n-G{\'o}mez, Gompper, and
			Winkler}}]{martin2018active}
	\bibinfo{author}{\bibfnamefont{A.}~\bibnamefont{Mart{\'\i}n-G{\'o}mez}},
	\bibinfo{author}{\bibfnamefont{G.}~\bibnamefont{Gompper}}, \bibnamefont{and}
	\bibinfo{author}{\bibfnamefont{R.}~\bibnamefont{Winkler}},
	\bibinfo{journal}{Polymers} \textbf{\bibinfo{volume}{10}},
	\bibinfo{pages}{837} (\bibinfo{year}{2018}).
	
	\bibitem[{\citenamefont{Nili}(2018)}]{Nili2018}
	\bibinfo{author}{\bibfnamefont{H.}~\bibnamefont{Nili}},
	\bibinfo{journal}{Scientific Reports} \textbf{\bibinfo{volume}{8}},
	\bibinfo{pages}{8328} (\bibinfo{year}{2018}).
	
	\bibitem[{\citenamefont{Mathijssen et~al.}(2016)\citenamefont{Mathijssen,
			Shendruk, Yeomans, and Doostmohammadi}}]{Mathijssen:PRL2016}
	\bibinfo{author}{\bibfnamefont{A.~J. T.~M.} \bibnamefont{Mathijssen}},
	\bibinfo{author}{\bibfnamefont{T.~N.} \bibnamefont{Shendruk}},
	\bibinfo{author}{\bibfnamefont{J.~M.} \bibnamefont{Yeomans}},
	\bibnamefont{and}
	\bibinfo{author}{\bibfnamefont{A.}~\bibnamefont{Doostmohammadi}},
	\bibinfo{journal}{Phys. Rev. Lett.} \textbf{\bibinfo{volume}{116}},
	\bibinfo{pages}{028104} (\bibinfo{year}{2016}).
	
	\bibitem[{\citenamefont{Malgaretti and Stark}(2017)}]{Malgarettijcp:2017}
	\bibinfo{author}{\bibfnamefont{P.}~\bibnamefont{Malgaretti}} \bibnamefont{and}
	\bibinfo{author}{\bibfnamefont{H.}~\bibnamefont{Stark}},
	\bibinfo{journal}{The Journal of Chemical Physics}
	\textbf{\bibinfo{volume}{146}}, \bibinfo{pages}{174901}
	(\bibinfo{year}{2017}).
	
	\bibitem[{\citenamefont{Pedley and Kessler}(1992)}]{pedley1992hydrodynamic}
	\bibinfo{author}{\bibfnamefont{T.}~\bibnamefont{Pedley}} \bibnamefont{and}
	\bibinfo{author}{\bibfnamefont{J.}~\bibnamefont{Kessler}},
	\bibinfo{journal}{Annual Review of Fluid Mechanics}
	\textbf{\bibinfo{volume}{24}}, \bibinfo{pages}{313} (\bibinfo{year}{1992}).
	
	\bibitem[{\citenamefont{Berke et~al.}(2008{\natexlab{b}})\citenamefont{Berke,
			Turner, Berg, and Lauga}}]{berke2008hydrodynamic}
	\bibinfo{author}{\bibfnamefont{A.~P.} \bibnamefont{Berke}},
	\bibinfo{author}{\bibfnamefont{L.}~\bibnamefont{Turner}},
	\bibinfo{author}{\bibfnamefont{H.~C.} \bibnamefont{Berg}}, \bibnamefont{and}
	\bibinfo{author}{\bibfnamefont{E.}~\bibnamefont{Lauga}},
	\bibinfo{journal}{Physical Review Letters} \textbf{\bibinfo{volume}{101}},
	\bibinfo{pages}{038102} (\bibinfo{year}{2008}{\natexlab{b}}).
	
	\bibitem[{\citenamefont{Malevanets and Kapral}(1999)}]{Malevanets:jcp:1999}
	\bibinfo{author}{\bibfnamefont{A.}~\bibnamefont{Malevanets}} \bibnamefont{and}
	\bibinfo{author}{\bibfnamefont{R.}~\bibnamefont{Kapral}},
	\bibinfo{journal}{The Journal of Chemical Physics}
	\textbf{\bibinfo{volume}{110}}, \bibinfo{pages}{8605} (\bibinfo{year}{1999}).
	
	\bibitem[{\citenamefont{Kapral}(2008)}]{Kapral:2008}
	\bibinfo{author}{\bibfnamefont{R.}~\bibnamefont{Kapral}},
	\emph{\bibinfo{title}{Multiparticle Collision Dynamics: Simulation of Complex
			Systems on Mesoscales}} (\bibinfo{publisher}{John Wiley and Sons, Inc.},
	\bibinfo{year}{2008}), pp. \bibinfo{pages}{89--146}, ISBN
	\bibinfo{isbn}{9780470371572}.
	
	\bibitem[{\citenamefont{Gompper et~al.}(2009)\citenamefont{Gompper, Ihle,
			Kroll, and Winkler}}]{Gompper2009}
	\bibinfo{author}{\bibfnamefont{G.}~\bibnamefont{Gompper}},
	\bibinfo{author}{\bibfnamefont{T.}~\bibnamefont{Ihle}},
	\bibinfo{author}{\bibfnamefont{D.~M.} \bibnamefont{Kroll}}, \bibnamefont{and}
	\bibinfo{author}{\bibfnamefont{R.~G.} \bibnamefont{Winkler}},
	\emph{\bibinfo{title}{Multi-Particle Collision Dynamics: A Particle-Based
			Mesoscale Simulation Approach to the Hydrodynamics of Complex Fluids}}
	(\bibinfo{publisher}{Springer Berlin Heidelberg}, \bibinfo{address}{Berlin,
		Heidelberg}, \bibinfo{year}{2009}), pp. \bibinfo{pages}{1--87}, ISBN
	\bibinfo{isbn}{978-3-540-87706-6}.
	
	\bibitem[{\citenamefont{Isele-Holder et~al.}(2015)\citenamefont{Isele-Holder,
			Elgeti, and Gompper}}]{isele2015self}
	\bibinfo{author}{\bibfnamefont{R.~E.} \bibnamefont{Isele-Holder}},
	\bibinfo{author}{\bibfnamefont{J.}~\bibnamefont{Elgeti}}, \bibnamefont{and}
	\bibinfo{author}{\bibfnamefont{G.}~\bibnamefont{Gompper}},
	\bibinfo{journal}{Soft matter} \textbf{\bibinfo{volume}{11}},
	\bibinfo{pages}{7181} (\bibinfo{year}{2015}).
	
	\bibitem[{\citenamefont{Anand and Singh}(2018)}]{anand2018structure}
	\bibinfo{author}{\bibfnamefont{S.~K.} \bibnamefont{Anand}} \bibnamefont{and}
	\bibinfo{author}{\bibfnamefont{S.~P.} \bibnamefont{Singh}},
	\bibinfo{journal}{Physical Review E} \textbf{\bibinfo{volume}{98}},
	\bibinfo{pages}{042501} (\bibinfo{year}{2018}).
	
	\bibitem[{\citenamefont{Malevanets and Kapral}(2000)}]{Kapral:jcp2000}
	\bibinfo{author}{\bibfnamefont{A.}~\bibnamefont{Malevanets}} \bibnamefont{and}
	\bibinfo{author}{\bibfnamefont{R.}~\bibnamefont{Kapral}},
	\bibinfo{journal}{The Journal of Chemical Physics}
	\textbf{\bibinfo{volume}{112}}, \bibinfo{pages}{7260} (\bibinfo{year}{2000}).
	
	\bibitem[{\citenamefont{Ihle and Kroll}(2001)}]{Ihle:PRE:2001}
	\bibinfo{author}{\bibfnamefont{T.}~\bibnamefont{Ihle}} \bibnamefont{and}
	\bibinfo{author}{\bibfnamefont{D.~M.} \bibnamefont{Kroll}},
	\bibinfo{journal}{Phys. Rev. E} \textbf{\bibinfo{volume}{63}},
	\bibinfo{pages}{020201} (\bibinfo{year}{2001}).
	
	\bibitem[{\citenamefont{Lamura et~al.}(2001)\citenamefont{Lamura, Gompper,
			Ihle, and Kroll}}]{lamura2001multi}
	\bibinfo{author}{\bibfnamefont{A.}~\bibnamefont{Lamura}},
	\bibinfo{author}{\bibfnamefont{G.}~\bibnamefont{Gompper}},
	\bibinfo{author}{\bibfnamefont{T.}~\bibnamefont{Ihle}}, \bibnamefont{and}
	\bibinfo{author}{\bibfnamefont{D.}~\bibnamefont{Kroll}},
	\bibinfo{journal}{EPL (Europhysics Letters)} \textbf{\bibinfo{volume}{56}},
	\bibinfo{pages}{319} (\bibinfo{year}{2001}).
	
	\bibitem[{\citenamefont{Lamura and Gompper}(2002)}]{lamura2002numerical}
	\bibinfo{author}{\bibfnamefont{A.}~\bibnamefont{Lamura}} \bibnamefont{and}
	\bibinfo{author}{\bibfnamefont{G.}~\bibnamefont{Gompper}},
	\bibinfo{journal}{The European Physical Journal E}
	\textbf{\bibinfo{volume}{9}}, \bibinfo{pages}{477} (\bibinfo{year}{2002}).
	
	\bibitem[{\citenamefont{Singh and Muthukumar}(2014)}]{Singh_2014_JCP}
	\bibinfo{author}{\bibfnamefont{S.~P.} \bibnamefont{Singh}} \bibnamefont{and}
	\bibinfo{author}{\bibfnamefont{M.}~\bibnamefont{Muthukumar}},
	\bibinfo{journal}{The Journal of chemical physics}
	\textbf{\bibinfo{volume}{141}}, \bibinfo{pages}{09B610\_1}
	(\bibinfo{year}{2014}).
	
	\bibitem[{\citenamefont{Malevanets and Yeomans}(2000)}]{Malevanets:EPL:2000}
	\bibinfo{author}{\bibfnamefont{A.}~\bibnamefont{Malevanets}} \bibnamefont{and}
	\bibinfo{author}{\bibfnamefont{J.~M.} \bibnamefont{Yeomans}},
	\bibinfo{journal}{EPL (Europhysics Letters)} \textbf{\bibinfo{volume}{52}},
	\bibinfo{pages}{231} (\bibinfo{year}{2000}).
	
	\bibitem[{\citenamefont{Ripoll et~al.}(2004)\citenamefont{Ripoll, Mussawisade,
			Winkler, and Gompper}}]{Ripoll:EPL:2004}
	\bibinfo{author}{\bibfnamefont{M.}~\bibnamefont{Ripoll}},
	\bibinfo{author}{\bibfnamefont{K.}~\bibnamefont{Mussawisade}},
	\bibinfo{author}{\bibfnamefont{R.~G.} \bibnamefont{Winkler}},
	\bibnamefont{and} \bibinfo{author}{\bibfnamefont{G.}~\bibnamefont{Gompper}},
	\bibinfo{journal}{EPL (Europhysics Letters)} \textbf{\bibinfo{volume}{68}},
	\bibinfo{pages}{106} (\bibinfo{year}{2004}).
	
	\bibitem[{\citenamefont{Huang et~al.}(2010{\natexlab{a}})\citenamefont{Huang,
			Chatterji, Sutmann, Gompper, and Winkler}}]{CCHuang2010}
	\bibinfo{author}{\bibfnamefont{C.}~\bibnamefont{Huang}},
	\bibinfo{author}{\bibfnamefont{A.}~\bibnamefont{Chatterji}},
	\bibinfo{author}{\bibfnamefont{G.}~\bibnamefont{Sutmann}},
	\bibinfo{author}{\bibfnamefont{G.}~\bibnamefont{Gompper}}, \bibnamefont{and}
	\bibinfo{author}{\bibfnamefont{R.}~\bibnamefont{Winkler}},
	\bibinfo{journal}{Journal of Computational Physics}
	\textbf{\bibinfo{volume}{229}}, \bibinfo{pages}{168 }
	(\bibinfo{year}{2010}{\natexlab{a}}), ISSN \bibinfo{issn}{0021-9991}.
	
	\bibitem[{\citenamefont{Huang et~al.}(2015)\citenamefont{Huang, Varghese,
			Gompper, and Winkler}}]{CCHuang2015}
	\bibinfo{author}{\bibfnamefont{C.-C.} \bibnamefont{Huang}},
	\bibinfo{author}{\bibfnamefont{A.}~\bibnamefont{Varghese}},
	\bibinfo{author}{\bibfnamefont{G.}~\bibnamefont{Gompper}}, \bibnamefont{and}
	\bibinfo{author}{\bibfnamefont{R.~G.} \bibnamefont{Winkler}},
	\bibinfo{journal}{Phys. Rev. E} \textbf{\bibinfo{volume}{91}},
	\bibinfo{pages}{013310} (\bibinfo{year}{2015}).
	
	\bibitem[{\citenamefont{Ihle and Kroll}(2003)}]{kroll:pre2003}
	\bibinfo{author}{\bibfnamefont{T.}~\bibnamefont{Ihle}} \bibnamefont{and}
	\bibinfo{author}{\bibfnamefont{D.~M.} \bibnamefont{Kroll}},
	\bibinfo{journal}{Phys. Rev. E} \textbf{\bibinfo{volume}{67}},
	\bibinfo{pages}{066705} (\bibinfo{year}{2003}).
	
	\bibitem[{\citenamefont{Winkler and Huang}(2009)}]{Winkler:jcp:2009}
	\bibinfo{author}{\bibfnamefont{R.~G.} \bibnamefont{Winkler}} \bibnamefont{and}
	\bibinfo{author}{\bibfnamefont{C.-C.} \bibnamefont{Huang}},
	\bibinfo{journal}{The Journal of Chemical Physics}
	\textbf{\bibinfo{volume}{130}}, \bibinfo{pages}{074907}
	(\bibinfo{year}{2009}).
	
	\bibitem[{\citenamefont{Whitmer and Luijten}(2010)}]{Whitmer:2010}
	\bibinfo{author}{\bibfnamefont{J.~K.} \bibnamefont{Whitmer}} \bibnamefont{and}
	\bibinfo{author}{\bibfnamefont{E.}~\bibnamefont{Luijten}},
	\bibinfo{journal}{Journal of Physics: Condensed Matter}
	\textbf{\bibinfo{volume}{22}}, \bibinfo{pages}{104106}
	(\bibinfo{year}{2010}).
	
	\bibitem[{\citenamefont{de~Gennes and Prost}(1995)}]{prost1995physics}
	\bibinfo{author}{\bibfnamefont{P.-G.} \bibnamefont{de~Gennes}}
	\bibnamefont{and} \bibinfo{author}{\bibfnamefont{J.}~\bibnamefont{Prost}},
	\emph{\bibinfo{title}{The physics of liquid crystals}},
	vol.~\bibinfo{volume}{83} (\bibinfo{publisher}{Oxford university press},
	\bibinfo{year}{1995}).
	
	\bibitem[{\citenamefont{Li and Tang}(2009{\natexlab{b}})}]{li2009accumulation}
	\bibinfo{author}{\bibfnamefont{G.}~\bibnamefont{Li}} \bibnamefont{and}
	\bibinfo{author}{\bibfnamefont{J.~X.} \bibnamefont{Tang}},
	\bibinfo{journal}{Physical review letters} \textbf{\bibinfo{volume}{103}},
	\bibinfo{pages}{078101} (\bibinfo{year}{2009}{\natexlab{b}}).
	
	\bibitem[{\citenamefont{Elgeti and
			Gompper}(2013{\natexlab{b}})}]{elgeti2013wall}
	\bibinfo{author}{\bibfnamefont{J.}~\bibnamefont{Elgeti}} \bibnamefont{and}
	\bibinfo{author}{\bibfnamefont{G.}~\bibnamefont{Gompper}},
	\bibinfo{journal}{EPL (Europhysics Letters)} \textbf{\bibinfo{volume}{101}},
	\bibinfo{pages}{48003} (\bibinfo{year}{2013}{\natexlab{b}}).
	
	\bibitem[{\citenamefont{Winkler}(2006)}]{winkler2006semiflexible}
	\bibinfo{author}{\bibfnamefont{R.~G.} \bibnamefont{Winkler}},
	\bibinfo{journal}{Physical review letters} \textbf{\bibinfo{volume}{97}},
	\bibinfo{pages}{128301} (\bibinfo{year}{2006}).
	
	\bibitem[{\citenamefont{Huang et~al.}(2010{\natexlab{b}})\citenamefont{Huang,
			Winkler, Sutmann, and Gompper}}]{Huang:Macromol2010}
	\bibinfo{author}{\bibfnamefont{C.-C.} \bibnamefont{Huang}},
	\bibinfo{author}{\bibfnamefont{R.~G.} \bibnamefont{Winkler}},
	\bibinfo{author}{\bibfnamefont{G.}~\bibnamefont{Sutmann}}, \bibnamefont{and}
	\bibinfo{author}{\bibfnamefont{G.}~\bibnamefont{Gompper}},
	\bibinfo{journal}{Macromolecules} \textbf{\bibinfo{volume}{43}},
	\bibinfo{pages}{10107} (\bibinfo{year}{2010}{\natexlab{b}}).
	
	\bibitem[{\citenamefont{Huang et~al.}(2012)\citenamefont{Huang, Gompper, and
			Winkler}}]{Chien-Cheng:2012}
	\bibinfo{author}{\bibfnamefont{C.-C.} \bibnamefont{Huang}},
	\bibinfo{author}{\bibfnamefont{G.}~\bibnamefont{Gompper}}, \bibnamefont{and}
	\bibinfo{author}{\bibfnamefont{R.~G.} \bibnamefont{Winkler}},
	\bibinfo{journal}{Journal of Physics: Condensed Matter}
	\textbf{\bibinfo{volume}{24}}, \bibinfo{pages}{284131}
	(\bibinfo{year}{2012}).
	
	\bibitem[{\citenamefont{Park and Butler}(2009)}]{park2009inhomogeneous}
	\bibinfo{author}{\bibfnamefont{J.}~\bibnamefont{Park}} \bibnamefont{and}
	\bibinfo{author}{\bibfnamefont{J.~E.} \bibnamefont{Butler}},
	\bibinfo{journal}{Journal of Fluid Mechanics} \textbf{\bibinfo{volume}{630}},
	\bibinfo{pages}{267} (\bibinfo{year}{2009}).
	
	\bibitem[{\citenamefont{Rahnama et~al.}(1995)\citenamefont{Rahnama, Koch, and
			Shaqfeh}}]{rahnama1995effect}
	\bibinfo{author}{\bibfnamefont{M.}~\bibnamefont{Rahnama}},
	\bibinfo{author}{\bibfnamefont{D.~L.} \bibnamefont{Koch}}, \bibnamefont{and}
	\bibinfo{author}{\bibfnamefont{E.~S.} \bibnamefont{Shaqfeh}},
	\bibinfo{journal}{Physics of Fluids} \textbf{\bibinfo{volume}{7}},
	\bibinfo{pages}{487} (\bibinfo{year}{1995}).
	
	\bibitem[{\citenamefont{Chen and Koch}(1996)}]{chen1996rheology}
	\bibinfo{author}{\bibfnamefont{S.~B.} \bibnamefont{Chen}} \bibnamefont{and}
	\bibinfo{author}{\bibfnamefont{D.~L.} \bibnamefont{Koch}},
	\bibinfo{journal}{Physics of Fluids} \textbf{\bibinfo{volume}{8}},
	\bibinfo{pages}{2792} (\bibinfo{year}{1996}).
	
	\bibitem[{\citenamefont{Leal and Hinch}(1971)}]{leal1971effect}
	\bibinfo{author}{\bibfnamefont{L.}~\bibnamefont{Leal}} \bibnamefont{and}
	\bibinfo{author}{\bibfnamefont{E.}~\bibnamefont{Hinch}},
	\bibinfo{journal}{Journal of Fluid Mechanics} \textbf{\bibinfo{volume}{46}},
	\bibinfo{pages}{685} (\bibinfo{year}{1971}).
	
	\bibitem[{\citenamefont{Kikuchi et~al.}(2002)\citenamefont{Kikuchi, Gent, and
			Yeomans}}]{kikuchi2002polymer}
	\bibinfo{author}{\bibfnamefont{N.}~\bibnamefont{Kikuchi}},
	\bibinfo{author}{\bibfnamefont{A.}~\bibnamefont{Gent}}, \bibnamefont{and}
	\bibinfo{author}{\bibfnamefont{J.}~\bibnamefont{Yeomans}},
	\bibinfo{journal}{The European Physical Journal E}
	\textbf{\bibinfo{volume}{9}}, \bibinfo{pages}{63} (\bibinfo{year}{2002}).
	
	\bibitem[{\citenamefont{Kikuchi et~al.}(2003)\citenamefont{Kikuchi, Pooley,
			Ryder, and Yeomans}}]{kikuchi2003transport}
	\bibinfo{author}{\bibfnamefont{N.}~\bibnamefont{Kikuchi}},
	\bibinfo{author}{\bibfnamefont{C.}~\bibnamefont{Pooley}},
	\bibinfo{author}{\bibfnamefont{J.}~\bibnamefont{Ryder}}, \bibnamefont{and}
	\bibinfo{author}{\bibfnamefont{J.}~\bibnamefont{Yeomans}},
	\bibinfo{journal}{The Journal of chemical physics}
	\textbf{\bibinfo{volume}{119}}, \bibinfo{pages}{6388} (\bibinfo{year}{2003}).
	
	\bibitem[{\citenamefont{Ripoll et~al.}(2007)\citenamefont{Ripoll, Winkler, and
			Gompper}}]{ripoll2007hydrodynamic}
	\bibinfo{author}{\bibfnamefont{M.}~\bibnamefont{Ripoll}},
	\bibinfo{author}{\bibfnamefont{R.}~\bibnamefont{Winkler}}, \bibnamefont{and}
	\bibinfo{author}{\bibfnamefont{G.}~\bibnamefont{Gompper}},
	\bibinfo{journal}{The European Physical Journal E}
	\textbf{\bibinfo{volume}{23}}, \bibinfo{pages}{349} (\bibinfo{year}{2007}).
	
	\bibitem[{\citenamefont{Padding and Briels}(2010)}]{padding2010translational}
	\bibinfo{author}{\bibfnamefont{J.}~\bibnamefont{Padding}} \bibnamefont{and}
	\bibinfo{author}{\bibfnamefont{W.~J.} \bibnamefont{Briels}},
	\bibinfo{journal}{The Journal of chemical physics}
	\textbf{\bibinfo{volume}{132}}, \bibinfo{pages}{054511}
	(\bibinfo{year}{2010}).
	
	\bibitem[{\citenamefont{Doi and Edwards}(1988)}]{doi1988theory}
	\bibinfo{author}{\bibfnamefont{M.}~\bibnamefont{Doi}} \bibnamefont{and}
	\bibinfo{author}{\bibfnamefont{S.~F.} \bibnamefont{Edwards}},
	\emph{\bibinfo{title}{The theory of polymer dynamics}},
	vol.~\bibinfo{volume}{73} (\bibinfo{publisher}{oxford university press},
	\bibinfo{year}{1988}).
	
	\bibitem[{\citenamefont{Singh et~al.}(2013)\citenamefont{Singh, Chatterji,
			Gompper, and Winkler}}]{singh2013dynamical}
	\bibinfo{author}{\bibfnamefont{S.~P.} \bibnamefont{Singh}},
	\bibinfo{author}{\bibfnamefont{A.}~\bibnamefont{Chatterji}},
	\bibinfo{author}{\bibfnamefont{G.}~\bibnamefont{Gompper}}, \bibnamefont{and}
	\bibinfo{author}{\bibfnamefont{R.~G.} \bibnamefont{Winkler}},
	\bibinfo{journal}{Macromolecules} \textbf{\bibinfo{volume}{46}},
	\bibinfo{pages}{8026} (\bibinfo{year}{2013}).
	
	\bibitem[{\citenamefont{Yang et~al.}(2015)\citenamefont{Yang, Theers, Hu,
			Gompper, Winkler, and Ripoll}}]{yang2015effect}
	\bibinfo{author}{\bibfnamefont{M.}~\bibnamefont{Yang}},
	\bibinfo{author}{\bibfnamefont{M.}~\bibnamefont{Theers}},
	\bibinfo{author}{\bibfnamefont{J.}~\bibnamefont{Hu}},
	\bibinfo{author}{\bibfnamefont{G.}~\bibnamefont{Gompper}},
	\bibinfo{author}{\bibfnamefont{R.~G.} \bibnamefont{Winkler}},
	\bibnamefont{and} \bibinfo{author}{\bibfnamefont{M.}~\bibnamefont{Ripoll}},
	\bibinfo{journal}{Physical Review E} \textbf{\bibinfo{volume}{92}},
	\bibinfo{pages}{013301} (\bibinfo{year}{2015}).
	
\end{thebibliography}

\end{document}